\documentclass[10pt,journal,compsoc]{IEEEtran}

\usepackage[normalem]{ulem}
\usepackage{fancyvrb,multirow,subfig}

\usepackage{xspace}
\usepackage{listings}
\usepackage{algorithm}
\usepackage{pifont}
\usepackage{amsmath}
\usepackage{fancyvrb}
\usepackage{url}
\usepackage{proof}
\usepackage{bussproofs}
\usepackage{tcolorbox}
\usepackage{balance}

\usepackage{algpseudocode}
\algrenewcommand\algorithmicrequire{\textbf{Input:}}
\algrenewcommand\algorithmicensure{\textbf{Output:}}

\newtheorem{property}{Property}

\newcommand{\SP}{\textsc{\em oo7}\xspace}

\begin{document}

\title{oo7: Low-overhead Defense against Spectre Attacks via Program~Analysis}
\author{\IEEEauthorblockN{Guanhua Wang,}
\IEEEauthorblockA{National University of Singapore}\\
\and
\IEEEauthorblockN{Sudipta Chattopadhyay,}
\IEEEauthorblockA{Singapore University of Technology and Design}\\
\and
\IEEEauthorblockN{Ivan Gotovchits,}
\IEEEauthorblockA{Carnegie Mellon University}\\
\and
\IEEEauthorblockN{Tulika Mitra \& Abhik Roychoudhury,}
\IEEEauthorblockA{National University of Singapore}\\
\thanks{To appear in IEEE Transactions on Software Engineering, 2020.}
}

\IEEEtitleabstractindextext{%
\begin{abstract}
The Spectre vulnerability in modern processors has been widely reported. The key insight in this vulnerability is that speculative execution in processors can be misused to access the secrets. Subsequently, even though the speculatively executed instructions are squashed, the secret may linger in micro-architectural states such as cache, and can potentially be accessed by an attacker via side channels. In this paper, we propose {\em oo7}, a static analysis approach that can mitigate Spectre attacks by detecting potentially vulnerable code snippets in program binaries and protecting them against the attack by patching them.
Our key contribution is to balance the concerns of effectiveness, analysis time and run-time overheads. We employ control flow extraction, taint analysis, and address analysis to detect
tainted conditional branches and speculative memory accesses. \SP can detect all fifteen purpose-built Spectre-vulnerable code patterns~\cite{spectremitigations}, whereas Microsoft compiler with Spectre mitigation option can only detect two of them. We also report the results of a large-scale study on applying \SP to over 500 program binaries (average binary size 261 KB) from different real-world projects.
We protect programs against Spectre attack by selectively inserting fences only at vulnerable conditional branches to prevent speculative execution. Our approach is experimentally observed to incur around 5.9\% performance overheads on SPECint benchmarks.
\end{abstract}}

\maketitle

\section{Introduction}
\label{sec:intro}
The Spectre~\cite{Kocher2018spectre} vulnerabilities in processors were revealed in early 2018. The attacks that exploit these vulnerabilities can potentially affect almost all modern processors irrespective of the vendor (Intel, AMD, ARM) and the computer system (desktop, laptop, mobile) as long as the processor performs speculative execution. 
Speculative execution~\cite{hennessy2011computer} is an indispensable micro-architectural optimizations for performance enhancement, ubiquitous in almost all modern processors except for the simplest micro-controllers. It is an aggressive optimization where the instructions are executed speculatively, but the temporary results created by the speculatively executed instructions are maintained in internal micro-architectural states that cannot be accessed by software. The results are committed to the programmer-visible architectural states (registers and memory) only when the speculation is found to be correct; otherwise, the internal micro-architectural states are flushed. The most common example is that of the conditional branches being predicted in hardware and the instructions along the predicted branch path are executed speculatively. Once the conditional branch direction is resolved, the instructions along the speculative path are squashed in case of wrong prediction. 


Spectre attacks exploit speculation to deliberately target the execution of certain ``transient" instructions. These transient instructions are speculatively executed, and are tricked to bring in secret data into the cache. These transient instructions are subsequently squashed but the secret remains, for example, in the cache. The attacker then carefully accesses the secret content (that is supposed to be hidden to the outside world) through different micro-architectural covert channels, for example, cache side-channel~\cite{yarom2014flush+}. 
The website~\cite{website} of Spectre states that ``As [Spectre] is not easy to fix, it will haunt us for a long time."

We focus on identifying program binaries that are vulnerable to Spectre attack and patch those binaries as a mitigation technique with minimal performance overhead . We present a comprehensive and scalable solution, called \SP, based on static program analysis.  Our solution employs control flow extraction, taint analysis and address analysis at the binary level. Moreover, {\em our analysis needs to model the transient instructions along the speculative path that has never been required in traditional program analysis dealing with only programmer visible execution. We have successfully introduced accurate modeling of speculative execution in \SP}.



Once vulnerable code snippets are detected by \SP, we introduce fence instructions at selected program points to prevent speculative execution and thereby protect the code from Spectre attack. We have validated the functional correctness of our protection mechanism with all fifteen litmus test codes from~\cite{spectremitigations} on Intel Xeon platform. We note that the current Spectre mitigation approach introduced by Microsoft C/C++ compiler~\cite{developerguidance}, detects and protects only 2 out of 15 litmus tests for Spectre vulnerabilities~\cite{spectremitigations}, whereas \SP can detect all fifteen purpose-built Spectre-vulnerable code patterns. We can launch successful Spectre attack to access arbitrary locations in the victim code prior to the insertion of fence insertions by \SP; but our attempts at Spectre attacks fail after \SP-directed automated identification and patching of the victim code. We experimentally measure the performance overheads from our selective fence insertion and find that the overheads are 5.9\% on average on SPECint benchmarks, thereby indicating the practicality of our approach.
We also report the results of a large-scale experimental study on applying \SP to over 500 program binaries (average binary size 261 KB) from different real-world projects.

We demonstrate that \SP can be tuned to defend against multiple different variants of Spectre attack (see TABLE~\ref{tb:list_all}) that exploit vulnerabilities in the victim code through speculative execution. We also note the limitations of our analysis-based approach in defending against certain variants of Spectre  attacks.
The variants that cannot be addressed by \SP have potential system-level solutions introduced by different vendors with reasonably low overhead~\cite{gruss2017kaslr, intelwhitepaper}. The Spectre variants handled by \SP with low performance overhead are either not amenable to system-level defense mechanisms, incur high performance overhead or escape detection with existing approaches. Thus \SP approach via binary analysis is complementary to all other efforts in mitigating the impact of security vulnerabilities due to speculative execution.




\subsection*{Contributions}

The contributions of this paper can be summarized as follows. First, we present a program analysis based approach called \SP for mitigating Spectre attacks. Our solution is based on binary analysis and does not involve changes to the underlying operating system and hardware. It uses taint analysis, address analysis and speculation modeling to check  potentially vulnerable program binaries, and inserts a small number of fences to mitigate the risks of Spectre attack. Our approach is accurate in identifying  all the litmus tests for Spectre vulnerabilities~\cite{spectremitigations}, has low performance overhead (average 5.9\% overhead for SPECint benchmark suite), and is scalable as evidenced by our analysis of over 500 large program binaries.

The main contribution of this work is in
	proposing and demonstrating an efficient static analysis
	based approach to accurately locate potential Spectre vulnerability
	in the code and then use well-established repair
	strategy (fences) to fix these selected vulnerable code fragments.
	We have successfully introduced accurate modeling
	of speculative execution in taint analysis to achieve this.
	The existing solutions cannot identify the vulnerable code
	fragments and hence repair all conditional branches to
	prevent speculative execution altogether resulting in significant
	performance overhead.

We show that our program analysis based approach can detect and mitigate certain
variants of Spectre vulnerabilities in the application code, but not all (see TABLE~\ref{tb:list_all}). Thus our
work provides an understanding of the class of attacks for which an analysis based mitigation may be suitable, and
for which a system level solution is suitable.

So far, no Spectre attack has been found in the wild.
We hope that the search for zero day Spectre attack in
the wild can be substantially accelerated via community
participation using our tool. Our tool is publicly available from 
{\tt \url{https://github.com/winter2020/oo7}}

\section{Spectre Variants}
\label{sec:overview}

\begin{center}
\begin{table*}[]
	\center
	\caption {The existing speculative execution based attacks and the ability of oo7 for handling them} \label{tb:list_all}
\begin{tabular}{l|lll}
	\hline \hline
	\textbf{Classification}                                                                                             & \textbf{Exploit name}          & \textbf{Public vulnerability name} & \textbf{oo7 capability}      \\ \hline
	\multirow{4}{*}{\textbf{Vulnerability in victim code}}                                                              & Spectre variant 1              & Bounds Check Bypass (BCB)          & Detect and patch vulnerable victim code \\
	& Spectre variant 1.1            & Bounds Check Bypass Store (BCBS)    & Detect and patch vulnerable victim code \\
	& Spectre variant 1.2            & Read-only protection bypass (RPB)  & Detect and patch vulnerable victim code \\
	& Spectre-NG variant 4           & Speculative Store Bypass (SSB)     & Potentially possible but not handled yet by oo7 \\ \hline
	\multirow{2}{*}{\textbf{BTB or RSB poisoning}}                                                                  & Spectre variant 2              & Branch Target Injection (BTJ)      & -                                \\
	& Spectre RSB                    & Return Mispredict                  & -                                \\ 
	\hline \hline
\end{tabular}
\end{table*}
\end{center}



A number of Spectre vulnerabilities that all take advantage of speculative execution in modern processors have been disclosed recently. A summary of these variants appear in TABLE~\ref{tb:list_all}. We classify the different vulnerabilities into three categories: 

\textit{(a) Vulnerability in victim code}: Many Spectre attacks rely on vulnerable code snippets inside the victim process and trigger speculative execution of the code snippet to read secret data by supplying carefully selected inputs to the victim process. We detect these vulnerabilities in \SP by  
identifying the potentially susceptible code fragments via binary analysis and then introducing fences at selected program points to prevent speculative execution and thereby harden the victim software against any such attacks. 

\textit{(b) BTB or RSB poisoning}: In these Spectre variants, the attacker poisons the Branch Target Buffer (BTB) or Return Stack Buffer (RSB) in the micro-architecture. The victim process, while using the poisoned BTB or the RSB for speculative execution, is then mislead to branch or return to a gadget that leaks the sensitive data. Any indirect branch or return instruction in the victim code is vulnerable to this attack and hence we do not attempt to mitigate these attacks in \SP. There exist potential solutions such as Retpoline~\cite{retpoline} or RSB refilling~\cite{koruyeh2018spectre} for these vulnerabilities. 

\textit{ (c) Transient out-of order execution}: These attacks can be directly launched by a malicious code (malware) without the 
requirement of any specific vulnerable code fragment or pattern in the victim process. Since the objective of \SP is to 
detect and repair vulnerable code fragments in general software, the detection of malware is orthogonal to the objective of \SP. 
Thus, we do not consider the detection of Spectre-style malicious code fragments in this work.

Unlike the first class of attacks where the 
defense mechanism is to harden the victim software, here \SP performs malware detection, i.e., it looks for malicious code patterns within a binary.  

\vspace*{0.1in}

\subsection{Vulnerability in Victim Code}
\label{sec:victim_code}

\paragraph*{\bf Spectre Variant 1} The following victim code fragment exhibits Spectre vulnerability {\tt Variant 1}.

\begin{Verbatim}[frame=single, fontsize=\footnotesize]
void victim_function_v01(size_t x) {
   if (x < array1_size) { //TB: Tainted Branch
      y = array1[x];  //RS: Read Secret y 		
      temp &= array2[y * 256]; //LS: Leak Secret y
   }
}
\end{Verbatim}
In this example, the parameter {\tt x} is under the attacker
control in the sense that {\tt x} can be influenced by external input. Hence we consider the conditional branch as a {\bf Tainted Branch (TB}).
The attacker first trains the branch predictor to expect that the branch will be true (i.e., the array bound check will pass). The attacker then invokes the code with an input {\tt x} value outside the bound of {\tt array1}. The branch predictor expects the branch condition to be true and the CPU
speculatively reads {\tt y} using malicious value {\tt x} outside the array bound. We call this action {\bf Read Secret (RS)} because {\tt y} can be a potential secret that is not legitimately accessible through malicious input without speculation.
This is followed by the CPU speculatively accessing {\tt array2} using an address that is dependent on the secret {\tt y} leading to cache state change. We call this action {\bf Leak Secret (LS)} because the change in the cache state lingers even after the CPU realizes that the branch prediction was wrong and squashes the speculatively executed instructions. The attacker can now launch cache side-channel attack~\cite{yarom2014flush+} to detect this change in cache state and discover the secret {\tt y}. Specifically, for {\tt Prime+Probe} side-channel attack, the attacker ensures that {\tt array2} was not cached before the memory access $\mathit{LS}$ by evicting the cache line through priming the cache set. 
Then the attacker triggers {\em LS} action to leak the secret to the cache side channel. Finally, the attacker performs the probe phase to get the timing of the memory accesses for  {\tt array2} and discovers the value of $y$. The multiplier {\tt 256} in {\tt array2[y*256]} guarantees that different values of {\tt y} lead to  different cache line access, and normally, this value is greater than or equal to the cache line size.


%

\paragraph*{\bf Spectre Variant 1.1} The idea behind the Spectre Variant 1.1,
also known as Bounds Check Bypass Store (BCBS), is to bypass bound
check and execute a store instruction speculatively~\cite{kiriansky2018speculative}.
In the following example, {\tt x} can potentially be under attacker control, hence, the
conditional {\tt x < array1\_size} is a {\em Tainted Branch (TB)}.
However, unlike the {\em Read Secret (RS)} in Spectre Variant 1, this
variant uses a {\bf Speculative Write (SW)} to modify arbitrary memory
location. For instance, the example modifies an arbitrary memory location
pointed to by {\tt array1[x]} when the conditional branch is mispredicted
for a value $x \geq array1\_size$. Although this speculative store is
squashed upon resolving the branch outcome, it can leak secret values
from the program. For instance, {\tt array1[x]} may overwrite the return
address and transfer control to a gadget that leaks arbitrary secret
value via a side-channel (similar to {\em LS} in Spectre Variant 1).

\begin{Verbatim}[frame=single, fontsize=\footnotesize]
void victim_function_v1.1(size_t x, y) {
  if (x < array1_size) { //TB: Tainted Branch
    array1[x] = y;  //SW: speculative Write
  }
}
\end{Verbatim}

\paragraph*{\bf Spectre Variant v1.2}
This vulnerability bypasses the protection enforced
by read-only memory, e.g., code pointers~\cite{kiriansky2018speculative}.
Consider the {\tt victim\_function\_v1.1} where the valuation of the
conditional captures whether {\tt x} points outside the read-only memory.
If {\tt x} is under attacker control, then the write to a read-only memory can
be speculatively executed and modify crucial data structures such as code
pointers in the cache. As a result, like Spectre Variant 1.1, the program
control may transfer to arbitrary location to execute attacker chosen
code. Like Spectre Variant 1.1,
this variant also requires the presence of {\em TB} and {\em SW}.

\paragraph*{\bf Spectre-NG Variant 4}
Spectre Variant 4, also called Speculative Store Bypass (SSB),
is based on the fact that the processor may execute a load instruction
speculatively even when a prior store instruction in program order is pending because the address for the store is not yet known. Thus a speculative load may read a stale value that should have been modified by a prior store instruction if they access the same memory address; in that case, the speculative load should be squashed after the store address is known. 

\SP can detect and patch victim binary code with potential Spectre variant 1, 1.1. and 1.2 vulnerabilities. \SP can potentially handle Spectre variant 4 by identifying the vulnerable code pattern but requires precise address analysis (that the load and the store are accessing the same memory address) that is not supported yet in our framework.

%
%

\subsection{BTB or RSB poisoning \protect\footnote{Even though these are considered as Spectre variants, they are very different from the vulnerability in victim code.}}

\paragraph*{\bf Spectre Variant 2} Most architectures support indirect branches
in the form of ``{\tt jmp [r1]}". For such jump instructions, the program
control is diverted to a location stored in the register {\tt r1}.
For improving program performance, the processor leverage Branch Target Buffer
(BTB) to store the frequently used target locations of branch instructions,
including indirect branches. An attacker can poison the Branch Target Buffer (BTB) to include
its preferred target 
locations. When the victim executes an indirect branch instruction, it consults this poisoned BTB and the speculative execution can potentially be misled to a target location chosen by the attacker. Any indirect branch is vulnerable to this attack. The indirect branches can be easily identified by
static analysis and mitigated by Retpoline~\cite{retpoline} approach. 


\paragraph*{\bf SpectreRSB}
SpectreRSB vulnerability~\cite{koruyeh2018spectre}
is similar to the Spectre Variant 2. Instead of poisoning the BTB
with attacker chosen location, the SpectreRSB vulnerability manipulates
the return stack buffer (RSB), which is used by the processor
to predict the return address. As a result of a successful exploit, a
function may return to an attacker controlled location due to the
mis-prediction of the return address inflicted by an attacker. Subsequently,
the program may execute arbitrary code in the attacker-controlled location
until the return address is finally resolved. All return instructions are potentially vulnerable to such exploit.
RSB refilling is a potential approach to mitigate SpectreRSB~\cite{koruyeh2018spectre}. 

Our proposed \SP approach focuses on identifying vulnerable code fragments (e.g. attacker 
controlled branch instructions) that can be fixed by the insertion of memory fence instructions. 
This is to prevent speculative execution with the objective to prevent the attack. Since Spectre 
Variant 2 and Spectre RSB need different solutions than patching fences, our \SP approach does not 
target fixing these subset of Spectre vulnerabilities.

\section{Related Work}
\label{sec:related}
\subsection{Mitigation in the future processors}
Intel has reportedly developed hardware fixes~\cite{Intel} in the form of improved process and privilege-level separation for {\em only} Spectre Variant 2. Three capabilities: Indirect Branch Restricted Speculation, Single Thread Indirect Branch Predictors, Indirect Branche Predictor Barrier will be supported in future products to mitigate the branch target injection attack. Vladimir et al.~\cite{kirianskydawg} proposed DAWG, which is a generic mechanism to isolate the cache side-channel by partitioning the cache ways to limit the data leakage across different secure domains. InvisiSpec~\cite{yan2018invisispec} is another new architecture design to defend  against Spectre-like attacks. InvisiSpec uses a Speculative Buffer (SB) to temporarily hold the data during speculative execution instead of directly loading the data to the cache. The data in SB will be finally visible to the cache hierarchy when the speculative load is safe to be committed. InvisiSpec slows down the execution by 21\%. Obviously, both DAWG and InvisiSpec cannot be used in the legacy systems.

\subsection{Mitigation in Legacy systems}
Several approaches have been applied to mitigate Spectre attacks in legacy system. Microsoft Visual C/C++ compiler~\cite{microsoftmitigation} provides a compiling option Qspectre to enable the mitigation of Spectre Variant 1 by inserting {\tt lfence} serializing instruction in the potential vulnerable code locations. However, the mitigation technique can detect only 2 out of the 15 litmus tests proposed by Paul Kocher~\cite{spectremitigations}. Speculative Load hardening~\cite{slharden} mitigates Spectre Variant 1 by inserting hardening instruction sequences that zeros out the pointers that have data dependency with the branch conditions. As it inserts hardening instructions at all conditional branches, the technique involves 36.4\% performance overhead. Oleksenko et al.~\cite{Oleksii} propose the introduction of artificial data dependencies to protect from Spectre attacks. This solution is coarse-grained and will effectively disable speculation between any conditional branch and subsequent load instructions. Moreover, the authors explicitly acknowledge the absence of precise taint tracking and leaves it to the developer to examine whether the potential vulnerable locations reported by the tool can be controlled by the attacker. Microsoft has developed Windows patches~\cite{msoft} through CPU microcode update for Spectre Variant 2 (but not for Variant 1). Moreover, this update has been reported to cause performance overhead (specially on older platforms) and system instability.  Google Chrome has developed ``Site isolation" mechanism that sandboxes the memory pages associated with each website to a separate process~\cite{chrome} at the cost of 10--13\% memory overhead. In contrast, \SP does not require either operating system or processor changes. Retpoline~\cite{turner2018retpoline} has been proposed for the gcc and LLVM compiler to mitigate Spectre version 2. Retpoline replaces vulnerable indirect branches with non-vulnerable instruction sequence that forces the CPU to jump to the real destination instead of the predicted target suggested by the BTB. Recently \cite{spectator} has proposed the use of symbolic execution and SMT solver for detecting Spectre Variant 1. This is a higher overhead and less scalable approach than ours. Moreover, no mitigation is proposed, as we do by inserting fence instructions.


Compared with all existing approaches for mitigation in the legacy system, \SP introduces the lowest performance overhead (around 5.9\% on SPECint) as \SP only hardens a small number of branches in repairing Spectre-like vulnerabilities.  Moreover, \SP is a flexible approach that can be tuned to defend against different variants of Spectre (as shown in this work) as well as detect malicious code fragments for Spectre-NG variant 3a and Spectre-NG LazyFP.

\section{Brief overview of our approach}

Our approach to identify vulnerable code fragments for Spectre variants 1, 1.1, 1.2 or malicious code fragments for Spectre-NG variant 3a and Spectre-NG LazyFP proceeds via static taint analysis of program binaries. {\em All input sources including files are initially
marked as tainted.} Taint propagation across instructions proceeds by usual computation of forward data
and control dependencies. Thus, for data dependency based taint propagation, if any of the operands of
an instruction is tainted, the result of the instruction is tainted. For control dependency based taint
propagation (also called as implicit flows in taint analysis literature), the decision taken by a branch, and hence the instructions conditionally executed owing to the decision are tainted; the identifiers (memory/registers) written by such tainted instructions are also treated as tainted. Details of the formal
treatment of taint propagation policies appear in Section \ref{sec:method}.

One of the novel aspects of our analysis is in considering speculative execution paths while capturing
taint propagation. Conceptually this is handled by considering both possibilities in a branch $b$, and checking
which instructions fall inside the speculative execution window of a branch meaning they can be speculatively
executed prior to a branch's outcome being known. There is no need to explicitly maintain the speculatively
executed paths as a separate set of bounded length paths, as long as we consider both directions of a branch in
our analysis. An instruction $i$ can be speculatively executed pending the outcome of a branch $b$, 
{\em only if the distance between $i$ and $b$ is less than the speculative execution window set by the 
processor}.

To check Spectre attack scenarios such as Spectre variant 1, we then need to look for a tainted branch instruction
({\em TB}), a load-like instruction which reads secret ({\em RS}) with the memory address read by RS being tainted,
and {\em RS} being potentially speculatively executed owing to {\em TB} being un-resolved. Note that once the occurrence of {\em TB} and {\em RS} are established, the secret data has already been speculatively accessed and can be potentially ex-filtrated via various side channels. To detect an instance of potential Spectre variant 1
vulnerability that is consummated via a cache side-channel attack, we need to also locate an instruction
{\em LS} to leak the secret, where the memory address accessed by $LS$ is dependent on the output of $RS$.
Once again, the detailed treatment of the condition for checking Spectre variant is deferred to the next section.

For forward taint propagation along all possible paths, we use the Binary analysis Platform (BAP) tool
\cite{brumley2011bap}. As BAP is based on conservative analysis, it can report false positives.
BAP leverages a set of 
techniques to construct the control flow graph for a binary. As is well known in binary analysis literature, accurate 
construction of control flow graph is a notoriously difficult problem owing to indirect branches. 
Tools like BAP use
forced execution \cite{peng2014x} and other techniques to construct a control flow graph. Forced execution (i.e., execution of both branch directions) is leveraged to construct all possible control flow edges of a branch. As the control flow paths are 
constructed, taint is propagated along the paths as per the taint sources and taint policies set in our approach (please see
next section for details). 

\paragraph*{On the challenges of binary analysis} During the translation of source code to the binary representation, optimizing compilers usually forfeit such properties as type preservation and control flow integrity, thus not only opening a possibility to a wide range of security vulnerabilities, such as buffer-overflow and control-flow hijacking but also complicating the task of reverse engineering and making sound static analysis of binary code merely impossible. The problem of control flow reconstruction is especially hard due to indirect branches where the 
branch targets are difficult to determine at compile-time. Therefore, any analysis of the binary code is doomed to be an approximation and a lot of care \cite{Maurer2018} should be taken to preserve the correctness of any analysis, and especially if that analysis involves a fixed point computation on the control flow graph, such as data flow analysis. In oo7 we leverage Primus, a Microexecution framework that tries to preserve the program behavior of the binary, while extracting control flow graph from the binary.

\paragraph*{On false negatives}
Our taint propagation policies and rules (see Section~\ref{sec:method}) avoid under-tainting, thus, in their own merit, they do 
not introduce any false negatives. Nonetheless, as our approach is based on the control flow graph constructed by the underlying tools, we cannot
guarantee zero false negatives, if the control flow graph constructed by BAP does not capture all possible flows. In other words,
the completeness of our taint analysis is modulo the completeness of the control
flow graph extraction in BAP. 
Also, in BAP, loops are unrolled to track the program dependencies across loop iterations and might be a source of false negatives 
in BAP analysis if the unrolling depth is low. However, with correctly provided loop bounds, this problem can be alleviated.

\paragraph*{On taint analysis and symbolic execution}
In identifying vulnerable code susceptible to many Spectre attack variants, the key
is to find attacker controlled branches, and memory locations that can be speculatively read/written pending outcome of attacker controlled branches. For this reason, we have employed static taint analysis and considered
read/write instructions that may be speculatively executed within the speculative execution window. It is possible to take a different, albeit higher-overhead approach. For example, instead of capturing the taint propagation along paths, one may summarize the execution behavior in a more fine grained fashion using symbolic execution, where the memory locations accessed by a read/write is captured as symbolic expressions over (tainted) input, instead of simply maintaining that the location accessed by a read/write is tainted. This leads to additional overhead of constraint accumulation and solving.
In this way, higher overhead detection approaches can be constructed
such as the recent work of \cite{spectator} that handles Spectre variant 1 and potentially variant 1.1.
This work~\cite{spectator} only reports results for different variants of the Spectre litmus tests with few lines of code.  As is
shown by our experiments, even a low overhead approach like ours consumes high analysis times on a few real-life programs from SPECint and OSS-Fuzz. Hence we feel higher overhead approaches are not scalable enough for possible real-life usage.  Furthermore, for any Spectre attack detection approach, it is not enough to detect leaks, the analysis needs to suggest a small number of fences to plug the leak. This is done in our approach. On the other hand, symbolic approaches like \cite{spectator} simply detect leaks without suggesting concrete fence instructions to harden Spectre vulnerable code.




\section{Spectre Vulnerability Detection}
\label{sec:method}

To describe our analysis, we use the notations in TABLE~\ref{symbol}.
We say that an instruction $inst$ is {\em tainted}, i.e., $\tau(inst)$ is true, if and only if
the instruction operates on some tainted operands. We first discuss
the checker for Spectre Variant 1.




\subsection{Detecting Spectre Variant 1}

An important concept that we need for our analysis is the {\em Speculative Execution Window}, abbreviated $\mathit{SEW}$.
We posit that information about $\mathit{SEW}$ needs to be exposed by processor designers for the sake of detecting Spectre
attacks. By default, it seems that $\mathit{SEW}$ can be set to the size of the re-order buffer in an out-of-order processor. However, if the size
of the re-order buffer is $n$, it is not sufficient to have a lookahead of $n$ instructions from a tainted
conditional branch $TB$, in our search for memory access $RS$, in order to detect Spectre attacks. For processor execution,
each instruction is decoded to a sequence of micro-ops. Each micro-op will occupy one slot of the re-order buffer during execution.
However, micro-ops can be fused \cite{opsfusion} both within an instruction as well as across instruction. When micro-ops are fused across instructions
(also called macro-fusion), the micro-ops of at most two instructions can be fused into a single micro-op. For this reason, if the size
of the re-order buffer is $n$, we conservatively set the Speculative Execution Window to $2n$ in our analysis, so as to avoid any false negatives in our analysis.

\begin{table}
\centering
\caption{Symbols used in describing \SP}
\label{symbol}
\resizebox{\linewidth}{!}{
\begin{tabular}{|c|p{5cm}|}
\hline
{\bf Symbol} & {\bf Interpretation}
\\ \hline\hline
$br(inst)$ & {\small $inst$ is a branch instruction}\\
\hline
$time(inst)$ & {\small $inst$ is an timing instruction, e.g., {\tt rdtsc}}\\
\hline
$mem(inst)$ & {\small $inst$ is a memory access instruction}\\
\hline
$load(inst)$ & {\small $inst$ is a load instruction}\\
\hline
$addr(inst)$ & {\small the  data memory address accessed by a memory-related  instruction $inst$} \\
\hline
$reg(inst)$ & {\small set of registers accessed by instruction $inst$} \\
\hline
\hline
$set(addr)$ & {\small cache set accessed by memory address $addr$} \\
\hline
$\tau(inst)$ & {\small instruction $inst$ is tainted} \\
\hline
$\tau(x)$ & {\small instruction operand $x$ is tainted where $x$ could be a register, memory
location or the value located in a memory location}\\
\hline
$\Delta(inst1,inst2)$ & {\small minimum no. of instructions executed to reach $inst2$ from $inst1$.
If $inst2$ is unreachable from $inst1$, then $\Delta(inst1,inst2) = \infty$.}\\
\hline
$\mathit{Dep}(inst, x)$ & {\small $x$ is data-dependent on instruction
$inst$}\\
\hline
$\mathit{CDep}(inst)$ & {\small set of instructions control-dependent on
$inst$}\\
\hline
$\mathit{val}(x)$ & {\small value located at memory address $x$}\\
\hline
$\mathit{SEW}$ & {\small Speculative Execution Window = $2n$, where $n$ is the size of re-order buffer in the processor}\\
\hline
\end{tabular}}
\end{table}

{\scriptsize
\begin{equation}
\boxed{
\label{eq:spectre-check}
\begin{split}
\Phi_{spectre} \equiv
br(\mathit{TB}) \wedge load(\mathit{RS}) \wedge mem(\mathit{LS}) \wedge
\\
\tau(\mathit{TB}) \wedge \tau(\mathit{addr(RS)}) \wedge \mathit{\tau(addr(LS))} \wedge
\\
\mathit{Dep}(\mathit{RS}, \mathit{addr(LS)}) \wedge
\left ( \Delta(\mathit{TB},\mathit{RS}) \leq \mathit{SEW} \right ) \wedge
\left ( \Delta(\mathit{TB},\mathit{LS}) \leq \mathit{SEW} \right )
\end{split}}
\end{equation}}

We now elaborate the checking condition for detecting Spectre.
\SP locates  $\mathit{TB}$, $\mathit{RS}$ and $\mathit{LS}$
by checking $\Phi_{spectre}$.
Intuitively, the first two lines of $\Phi_{spectre}$ capture the presence
of tainted branch instructions $\mathit{TB}$ and tainted memory-access
instructions $\mathit{RS}$ and $\mathit{LS}$. The last line shows
that $\mathit{RS}$ and $\mathit{LS}$ are located within the speculation
window of  $\mathit{TB}$, and they are data-dependent.
$\Phi_{spectre}$ reflects Spectre variant 1. Later we show
that $\Phi_{spectre}$ can easily be modified to detect other
Spectre variants, for example, variant 1.1.


\subsection{Taint Analysis}
\label{sec:taint}
We use taint analysis~\cite{brumley2011bap} to determine whether conditional branch
instructions (e.g., $\mathit{TB}$) and the memory-access instructions
(e.g., $\mathit{RS}$ and $\mathit{LS}$) can be controlled via untrusted
inputs. In the following, we outline the taint propagation policies and
rules used to detect Spectre vulnerabilities.
%
%
%
To illustrate our taint propagation policy, we use the following
instructions.
\begin{itemize}
\item
$z = x\ {\tt op} \ y$ : Binary operation on register $x$ and register $y$.
The operation {\tt op} can be either arithmetic operation (e.g., addition or subtraction)
or a logical operation (e.g., a logical comparison).

\item $y = {\tt op}\ x$ : Unary operation on register $x$. The operation
{\tt op} can either be arithmetic (e.g., unary minus) or a logical one
(e.g., logical negation).

\item $y = {\tt load}(x)$ : Loads value from memory address $x$ to register $y$.

\item $y = {\tt store}(x)$ : Stores value from register $x$ to memory address $y$.

\item ${\tt branch}(L,x)$ : Branch to label $L$ if the logical formula $x$ is {\em true}.
\end{itemize}

\subsubsection*{Taint Propagation Policies:}

Initially, all variables that read value from
un-trusted sources ({\em e.g.}, files, network) are tainted. The taints from these
variables are then propagated via a well-defined set of rules shown in the following;
for each rule, the premises appear on the top of the horizontal bar and the conclusions
appear below the horizontal bar.
Our taint propagation tracks both data dependencies and control dependencies
(also known as implicit flows in taint analysis).
 Typically such implicit flows come in the form of the tainted data enabling or disabling
a branch condition $b$, and the outcome of $b$  affecting the computation
of a variable that would not be tainted otherwise purely by tracking of data dependencies.

{\footnotesize

\begin{tcolorbox}[title=Taint Propagation Rules, colback=white!25,arc=0pt,auto outer arc]
\begin{prooftree}
\AxiomC{$z = x\ {\tt op}\ y$}
\AxiomC{$\tau(x) \vee \tau(y)$}
\LeftLabel{[Binary operation]}
\BinaryInfC{$\tau(z) \wedge \tau(inst)$}
\end{prooftree}
\begin{prooftree}
\AxiomC{$y = {\tt op}\ x$}
\AxiomC{$\tau(x)$}
\LeftLabel{[Unary operation]}
\BinaryInfC{$\tau(y) \wedge \tau(inst)$}
\end{prooftree}
\begin{prooftree}
\AxiomC{$y = {\tt load}(x)$}
\AxiomC{$\tau \left ( val(x) \right )$}
\LeftLabel{[Memory load]}
\BinaryInfC{$\tau(y) \wedge \tau(inst)$}
\end{prooftree}
\begin{prooftree}
	\AxiomC{$y = {\tt load}(x)$}
	\AxiomC{$\tau  (x)  $}
	\LeftLabel{[Memory load]}
	\BinaryInfC{$\tau(y) \wedge \tau(inst)$}
\end{prooftree}
\begin{prooftree}
	\AxiomC{$y = {\tt store}(x)$}
	\AxiomC{$\tau (x)$}
	\LeftLabel{[Memory store]}
	\BinaryInfC{$\tau \left ( val(y) \right ) \wedge \tau(inst)$}
\end{prooftree}

\begin{prooftree}
\AxiomC{$br(inst)$}
\AxiomC{$\tau (x)$}
\LeftLabel{ [${\tt branch}(L,x)$]}
\BinaryInfC{$\tau(inst)$}
\end{prooftree}
\begin{prooftree}
\AxiomC{$\tau (x)$}
\AxiomC{$\mathit{Tinst}=\mathit{CDep}(inst)$}
\LeftLabel{ [${\tt branch}(L,x)$]}
\BinaryInfC{$\forall t \in \mathit{Tinst}.\ \tau(t) \wedge \forall w \in \mathit{Write(Tinst)}.\ \tau(w)$}
\end{prooftree}
\end{tcolorbox}
}

\subsubsection*{Taint Propagation Rules:}
Based on the discussions in the preceding paragraphs, the taint propagation
rules are shown. In the taint propagation rules, we assume that $inst$
captures the current instruction for which the taint propagation is being
computed. $\mathit{Write(Tinst)}$ captures the set of operands written by
a set of instructions $\mathit{Tinst}$. In the following, we discuss a few
salient features of our taint propagation rules:

\smallskip\noindent
\textbf{Computation:} For instructions involving computations (e.g., addition,
subtraction and multiplication), we consider them tainted if and only if they
operate on tainted operands. Moreover, the result computed by such tainted
instructions are also marked tainted. This is because the outcome of these
instructions can be controlled by an attacker if they operate on tainted
operands. As \SP operates at the binary code level, most computations can be
captured via either binary or unary operations, as shown in our taint
propagation rules.

\smallskip\noindent
\textbf{Memory Load:}
Accounting taint propagation through memory load instruction is crucial
for the effectiveness of \SP. To this end, we need to consider two different
scenarios: {\em (i)} taint propagation through the value being loaded, and
{\em (ii)} taint propagation through the accessed memory address for the
load. In both cases, the loaded register can be controlled by an attacker,
as either the accessed address or the value located therein can be
manipulated to load an attacker controlled value. To model these taint
propagation rules, we need to accurately track the tainted status of the
accessed memory regions and this, in turn, involves a conservative analysis
of possible aliases in the program.

\smallskip\noindent
\textbf{Memory Store:} We note that a tainted value can be stored in an
arbitrary memory address $y$. For such an operation, we conclude that
the value located in address $y$ can be tainted, as captured by the
predicate $\tau(val(y))$. Such a store operation, however, does not conclude
anything about $\tau(y)$, as the store operation ``${\tt y = store(x)}$" cannot
control the address $y$ via the tainted operand $x$.

\smallskip\noindent
\textbf{Conditional Branch:}
Conditional branches are involved in accounting for both the explicit and implicit
(i.e., taints through control dependencies) propagation of taints. For example,
the outcome of a conditional branch can be controlled by an attacker
(i.e., tainted) if the variable $x$ in ${\tt branch}(L,x)$ is tainted. Moreover,
we discover the set of instructions $Tinst$ that are control dependent on a branch
instruction. If the branch is tainted, then the value of any variable written
by instructions in $Tinst$ is indirectly (i.e., implicitly) controlled by the
attacker. Consequently, we mark all such values tainted, as captured by our
taint propagation rules.

For the sake of brevity, the aforementioned taint propagation rules are described
based on a simplified syntax of low-level binary code. The taint propagation rules
avoid any under-tainting. However, these taint propagation rules may lead
to over-tainting as described in the following.

\smallskip\noindent
\textbf{Sources of Over-tainting:}
Over-tainting in \SP may lead to some instruction $inst$ or data element (i.e., a
value or an address) $x$ to be tainted (i.e., $\tau(inst)$ or $\tau(x)$ holds,
respectively) even in the absence of any feasible execution where $\tau(inst)$
or $\tau(x)$ is true. Such a phenomenon may occur due to the following scenarios:

\begin{enumerate}
\item {\bf Conservative extraction of CFG:} It is often challenging to extract
an accurate control flow graph from the binary code. This is primarily 
due to the difficulty in precisely identifying the branch targets, such as the
targets of indirect branches and calls. This, in turn, leads to conservative
approximation of control flow edges in the extracted CFG. As our taint
propagation rules walk the CFG and leverage control dependency graph (CDG) to
compute implicit taint propagation, the method might lead to over-tainting due to the
additional edges in the CFG and CDG.

\item {\bf Conservative alias analysis:} As our taint propagation rules
involve taint tracking through both the memory addresses and the values in these
memory addresses, \SP may over-taint due to a conservative alias analysis.
For example, two memory addresses $x$ and $y$ might be considered aliases
even if $x$ and $y$ do not point to the same memory address in any feasible
execution. Nonetheless, our taint propagation rules will conservatively
assume that all values in memory locations pointed by $y$ are tainted, given
that the values in memory locations pointed by $x$ can be attacker controlled.

\item {\bf Approximate memory model:} An accurate taint tracking requires
the tainted status of each possible memory address that can be accessed by
a program. This might often be expensive and the static analysis may resort
to conservative approximation for scalability. For instance,
even if a particular data element in an aggregate (e.g., a structure) is
tainted, to speed up the analysis, \SP might conservatively assume values
in all the memory addresses occupied by the aggregate data structure to be
tainted.

\item {\bf Non-analyzable function calls:} The return values from non-analyzable
function calls are conservatively considered to be tainted. These function
calls might be libraries or other third-party software that cannot
be analyzed.

\end{enumerate}

\smallskip\noindent
\textbf{Properties:}
In summary, \SP guarantees the following crucial property via its taint propagation rules:
\begin{property}
\label{prop:taint-rules}
{\em For a given program $P$, consider an arbitrary register or memory location $x_P$
that can be controlled by an attacker. Moreover, let the outcome of instruction
$inst_P$ may also be controlled by an attacker. The taint propagation rules of
\SP guarantee that $\tau(x_P)$ and $\tau(inst_P)$ hold. }
\end{property}

We consider all inputs arriving from external sources (e.g., network, files, command
lines) are tainted initially. These are called taint sources. Our taint propagation
rules then avoid any under-tainting, by considering the forward transitive closure
of all control and data dependencies from the taint sources.

%
%

\subsection{Detecting other Spectre variants}
\label{sec:spectre-detect-variant}

In the preceding section, we discussed the detection of
Spectre variant 1~\cite{Kocher2018spectre}. Note that Spectre variant 1 can leak the secret data in other ways instead of performing the $\mathit{LS}$ action. Such variants
can be detected via simple manipulation of $\Phi_{spectre}$:
{\scriptsize
\begin{equation}
\boxed{
\label{eq:spectre-check-weak}
\begin{split}
\Phi_{spectre}^{weak} \equiv
br(\mathit{TB}) \wedge load(\mathit{RS})\wedge
\tau(\mathit{TB}) \wedge \tau(\mathit{addr(RS)}) \wedge
\\
\left ( \Delta(\mathit{TB},\mathit{RS}) \leq \mathit{SEW} \right )
\end{split}}
\end{equation}}

Our \SP approach can be fine tuned to detect other Spectre variants.
For instance, consider Spectre Variant 1.1 (cf. TABLE~\ref{tb:list_all}). Such
a variant can easily be detected by the following condition:
{\scriptsize
\begin{equation}
\boxed{
\label{eq:spectre-check-v1.1}
\begin{split}
\Phi_{spectre}^{v1.1} \equiv
br(\mathit{TB}) \wedge store(\mathit{SW})\wedge
\tau(\mathit{TB}) \wedge \tau(\mathit{addr(SW)}) \wedge
\\
\left ( \Delta(\mathit{TB},\mathit{SW}) \leq \mathit{SEW} \right )
\end{split}}
\end{equation}}
where $store(\mathit{SW})$ captures the presence of a speculative write instruction,
as needed to exploit Spectre Variant 1.1. 

Spectre Variant 1.2 (read-only protection
bypass) needs exactly the same condition $\Phi_{spectre}^{v1.1}$ to be satisfied,
except that the speculative write ($\mathit{SW}$) happens to be in read-only memory.
For the rest of the paper, we do not distinguish between Spectre Variant 1.1 and 1.2,
as \SP uses the same condition to detect both the variants.

To detect Spectre Variant 4, we need to check whether a load instruction ($\mathit{RS}$)
follows a store ($\mathit{WR}$) to the same address, yet $\mathit{RS}$ can speculatively load a value not yet written
by $\mathit{WR}$.
Checking for this condition requires accurate address analysis, more accurate than what we can currently support.
We are currently working in this direction.

\begin{figure*} [ht]
	\centering
	\includegraphics[width=0.95\linewidth]{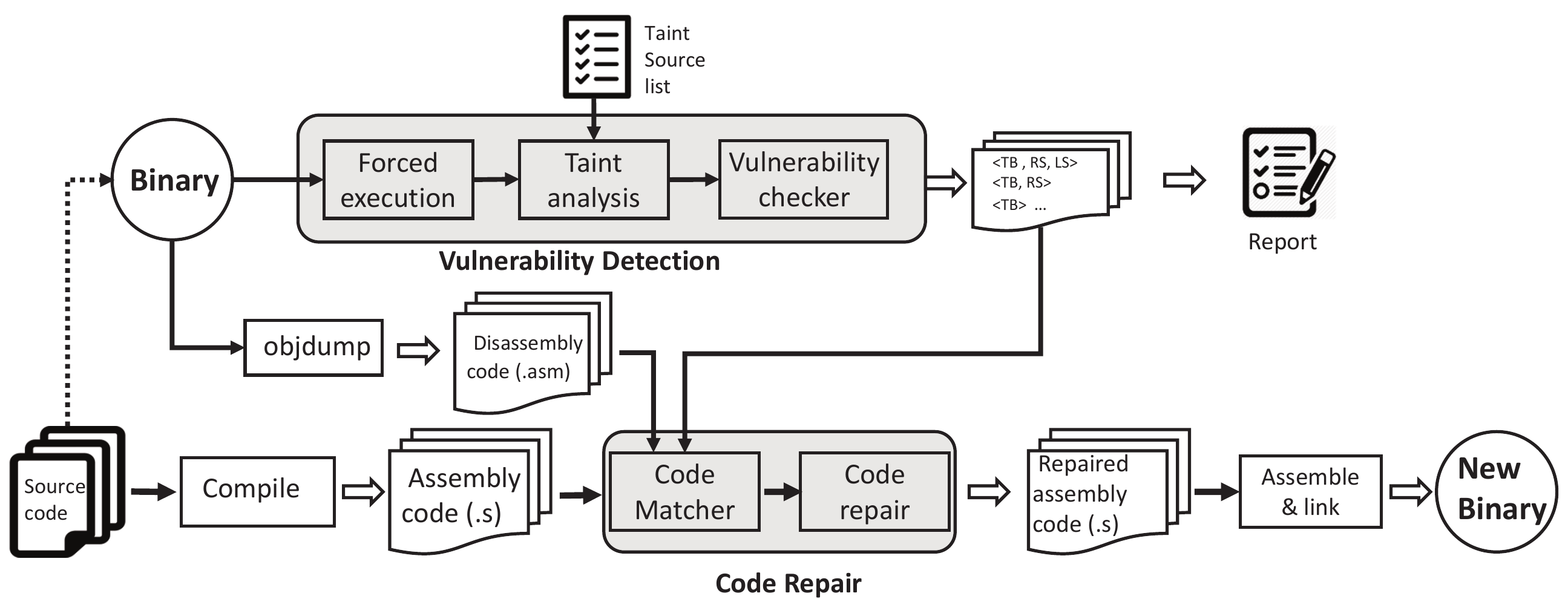}
	\caption{Overview of {\SP} framework. The components in grey represent the core modules of \SP: vulnerability detection module and code repair module.}
	\label{fig:framework}
\end{figure*}
\subsection{Code repair}
\label{sec:spectre-repair}

Our repair strategy is based on systematically inserting memory fences
after each tainted branch (i.e., $\mathit{TB}$) in vulnerable code fragments for Spectre variants
1, 1.1 and 1.2.
The original article describing Spectre attacks~\cite{Kocher2018spectre}
suggests insertion of memory fences following each conditional branch.
However, using our analysis, we can obtain the exact sequence
$\langle \mathit{TB}, \mathit{RS}, \mathit{LS} \rangle$ (for Variant~1)
or the sequence $\langle \mathit{TB}, \mathit{SW} \rangle$ (for
Variant~1.1 and Variant~1.2) vulnerable to Spectre attacks. As a result,
we can accurately locate the program point where the memory fence should
be inserted. In particular, we insert memory fences following $\mathit{TB}$
instruction and immediately before the execution of $\mathit{RS}$ and
$\mathit{SW}$, respectively, for Spectre Variant 1 and Variant 1.1, 1.2. This
prevents execution from loading the secret value into the cache
(for Variant 1) and writing to an attacker-controlled location (for
Variant 1.1, 1.2) speculatively.

Nevertheless, inserting memory fences may affect the overall program performance.
\SP inserts memory fences only for the branches identified as $\mathit{TB}$ (for
variants 1, 1.1 and 1.2). This has less overhead than inserting fences after each conditional branch, or after
each tainted conditional branch. We show empirically that such a strategy has
acceptable performance overheads of average 5.9\% for SPECint benchmarks. 
When selecting the fence instruction, we observed in experiments that {\tt cpuid}~\cite{cpuid}, {\tt lfence} and {\tt mfence} are able to prevent Spectre attacks (while {\tt sfence} cannot). However, if {\tt cpuid} is used for repairing Spectre vulnerability in the assembly code, then it modifies the general-purpose registers. 
Specifically, the return value of {\tt cpuid} is stored in registers such as {\tt EAX}, {\tt EBX}, {\tt ECX} or {\tt EDX}. Thus, additional instructions are required to store and 
restore the impacted register(s) before and after invoking {\tt cpuid}, which introduces additional performance overhead. On the other hand, {\tt lfence} is officially confirmed and recommended by Intel~\cite{intelwhitepaper} to mitigate Spectre attack. Therefore, in this work, we use only the {\tt lfence} instruction to repair vulnerable code.


\section{Implementation}
\label{sec:implementation}

Fig.~\ref{fig:framework} provides an overview of \SP tool. \SP contains
two main modules: a vulnerability detection module for detecting the Spectre
vulnerabilities, 
and a code repair module to fix the Spectre vulnerabilities.


\textbf{Vulnerability detection module.}
The vulnerability detection module of \SP is supported by three major technologies: forced execution, taint analysis and vulnerability checker.


 {\em Forced execution}~\cite{peng2014x, shoshitaishvili2016state}, as its name suggests, forces the program to execute along all possible paths by predicting the branch outcomes to both true and false. The capability of forced execution is in exploring the different execution paths and simulating the execution to expose the behavior of a given program. Forced execution satisfies the semantics of speculative execution, because speculative execution may lead the processor to execute the instructions on both outcomes of a branch when the branch prediction is wrong. Forced execution engine explores all possible paths by maintaining a pool of execution paths that may be explored in future by switching more predicates. A predicate is represented as a tuple $(I_{src}, I_{dst})$ where $I_{src}$ and $I_{dst}$ denote the source instruction and forced execution target (the branch outcomes), respectively. New predicates are added when a branch instruction is evaluated. As forced execution is a well-known technique, for details of forced execution technique, we refer the readers to relevant previous works~\cite{peng2014x, shoshitaishvili2016state,Kim:2017:JFE:3038912.3052674, xu2009constructing}.

{\em Taint analysis} performed by the taint propagation engine tracks the data and instruction that can be controlled by the attacker. The taint analysis works along with the forced execution engine. When forced execution engine evaluates a {\tt call} instruction, it checks if the destination of the {\tt call} is in the taint source list. The taint source list is a set of APIs which can import the data to the program from the un-trusted channels such as network, user input, file reader. In the implementation, we consider all interfaces (e.g. {\tt fgetc(), recv()}) in the commonly used libraries (e.g. {\tt glibc}) which import data from outside of program as the taint sources. If the destination of a {\tt call} instruction executed by the forced execution engine is in the taint source list, the taint propagation engine marks the imported data as a tainted object. After the tainted object is imported from the taint source, the taint engine propagates this tainted object during the execution by applying the taint propagation rules explained in section~\ref{sec:taint}.

{\em Vulnerability checker} detects whether the current state satisfies the conditions of arbitrary vulnerabilities presented in section~\ref{sec:spectre-detect-variant}. The vulnerability checker works after the taint engine taints an instruction. When the taint engine taints a new conditional branch instruction, the vulnerability checker records this tainted branch and sets up a Speculative Execution Windows (SEW) for it. SEW is decremented by one at the end of the evaluation of each instruction along the execution path. The vulnerability checker reviews whether a memory instruction satisfies the condition for $\mathit{RS}$ or  $\mathit{LS}$ before the SEW is decremented to zero. The vulnerability checker records down  $\mathit{TB}$, $\mathit{RS}$ and  $\mathit{LS}$ as a potentially vulnerable code fragment for the final report.

We adopt BAP~\cite{brumley2011bap} as our primary analysis platform. BAP provides a toolkit for implementing automated binary analysis and it supports multiple architectures such as x86, x86-64, ARM, PowerPC, and MIPS. BAP lifts binary code into RISC-like intermediate representation (IR) named BAP Instruction Language (BIL). Program analysis is performed using the BIL representation and it is architecture independent. BAP contains a microexecution framework named Primus performs forced execution and taint analysis. BAP provides several interfaces to export crucial information to other analysis modules during the analysis. We implement the vulnerability checker bases on these interfaces.

\textbf{Vulnerability repair module.}
Once a vulnerable code fragment is detected in the binary, we
locate the corresponding assembly code for repair. To this end,
we first mark the address(es) of Spectre vulnerable code, as
obtained during the detection stage of \SP.
Concurrently, we obtain the disassembled code from the binary
and the assembly code from the source (via ``{\tt -S}" option in
gcc compiler).
As most optimizations are employed during the
compilation stage rather than assembler stage, there is no substantial difference
between the assembly code and the respective disassembled code.
This allows us to easily map the disassembled code back to the
assembly code and locate the instructions vulnerable to Spectre attack.

Finally, our repair module directly modifies the assembly code by
inserting memory fence instructions in the appropriate location
(e.g., inserting {\tt lfence} before {\tt RS} or {\tt SW} for mitigating Spectre
Variants).

\section {Evaluation Setup}
In this section, we present the details of our evaluation setup. We first describe the programs used in our experiment. Then we introduce the platform used for the evaluation.
\subsection{Subject Programs}
We conduct evaluation on three sets of subject programs.
\begin{itemize}
\item
We first apply {\em oo7} on 15 code examples purpose-built to demonstrate different variations of Spectre vulnerabilities from Paul Kocher's blog post~\cite{spectremitigations}. We call these {\em Litmus Tests}.
\item Next, we conduct evaluation on SPECint benchmarks, which have been well-studied by the computer architecture community. These are detailed in TABLE~\ref{tb:program}. We concentrate on complete analysis of the SPECint (integer) benchmark suite because it includes more control-intensive code compared to SPECfp (floating point) and Spectre exploits vulnerability through conditional branches. SPECint benchmark suite contains 18.31\% branches in the instruction mix compared to only 5.75\% for SPECfp~\cite{kejariwal2008comparative}. 

\item
Last but not the least, we conduct evaluation with a large number of software projects from Google OSS-Fuzz repository~\cite{oss-fuzz} and GitHub. The program binaries in these project include the main application and miscellaneous support tools. TABLE~\ref{tb:benchmark} summarizes the characteristics of these projects consisting of a total of 507 program binaries with size ranging from 8.5KB to 21.8MB (average size 261.4KB).
\end{itemize}

\subsection{Evaluation Platform}
We conduct experimental evaluation on Intel Xeon Gold 6126~\cite{xeongold} running at 2.6GHz with 192GB memory. The underlying micro-architecture is Skylake with 224-entry reorder buffer (ROB)~\cite{skylake}. Due to the potential expansion of instructions into micro-operations and subsequent fusion in x86 micro-architectures (Section~\ref{sec:method}), we conservatively set the speculative window to twice the effective ROB size, i.e., $\mathit{SEW} = 448$.
Intel Xeon Gold 6126 is equipped with 12 cores and 19.25MB non-inclusive shared last-level cache (LLC) with 64 byte line size. The LLC cache miss penalty is about 200 cycles. Non-inclusive LLC is more secure than the inclusive cache and can thwart certain LLC based side-channel attacks (e.g., {\tt Flush+Flush, Prime+Probe}). However, it is still vulnerable to {\tt Flush+Reload} attack. Thus the Spectre and Meltdown attacks can potentially be carried out in this platform.

\section{Evaluation Results}
\label{sec:evaluation}

Our evaluation investigates three different aspects:
\begin{enumerate}
\item \textit{Effectiveness:} How effective is \SP in detecting Spectre vulnerabilities in program binaries?
\item \textit{Analysis Time:} How long is \SP analysis time to detect Spectre vulnerabilities?
\item \textit{Performance Overhead:} How much is the performance overhead introduced by \SP to protect vulnerable code fragments?
\end{enumerate}

\subsection{Evaluation on Litmus Tests}
\label{sec:specfix-effectiveness}

{\em oo7} can correctly identify all code snippets purpose-built with different variations of Spectre vulnerabilities~\cite{spectremitigations} as potential victim code fragments. 14 code examples are identified with taint propagation only along data dependencies. The remaining code example is detected with taint propagation along program (both control and data) dependencies.

The latest Microsoft Visual C++ compiler~\cite{developerguidance} has
integrated {\tt /Qspectre} switch for mitigating a limited set of potentially
vulnerable code patterns related to Spectre vulnerabilities. Specifically, after
compiling an application with {\tt /Qspectre} enabled, Visual C++ compiler
attempts to insert an {\tt lfence} instruction upon detecting Spectre code patterns.
Paul Kocher~\cite{spectremitigations} has evaluated the Microsoft compiler using 15 litmus tests.
The blog post \cite{spectremitigations} mentions
that only two of the micro-benchmarks are identified and protected
by Visual C++ compiler. In contrast, \SP can correctly detect all 15 code examples as potential victims.

The example ({\tt v13}~\cite{spectremitigations}) that requires taint propagation along both control and data dependencies is given below.
{
\begin{Verbatim} [frame=single, fontsize=\footnotesize]
__inline int is_x_safe(size_t x) {
  if (x < array1_size) return 1; else return 0;
}
void victim_function_v13(size_t x) {
  if (is_x_safe(x)) temp &= array2[array1[x]*512];
}
\end{Verbatim}
}

The branch in the victim function {\tt victim\_function\_v13} is tainted as the return
value of {\tt is\_x\_safe(x)} is controlled via untrusted input {\tt x}.
However, the return value  of {\tt is\_x\_safe(x)} is control-dependent and not data-dependent
on {\tt x}. Thus \SP can detect this code pattern as potential vulnerability only if both data- and control-dependent taint propagation are applied.

\subsection{Validation of Patching} 

We design an attacker process to steal secrets via cache side-channel from the victim process (litmus test example~\cite{spectremitigations}) once the secret data is brought into the cache through Spectre attack. We successfully extracted data from arbitrary memory locations in the victim process on our platform for 10 out of 15 litmus tests. We then allow \SP to automatically insert {\tt lfence} instructions at appropriate program locations to prevent speculation in vulnerable code fragments. We verify that the attacker process can no longer extract data from the victim processes running with the \SP fix for all 10 litmus tests even though we had successfully extracted secret data from all of them before patching. 

\begin{table}[]
	\caption{{\tt bzip2} randomly inserted with $\mathit{\sigma}$ vulnerable functions. VF\_source: number of functions inserted in the source code;  VF\_oo7: number of vulnerable functions identified by \SP. Analysis time: Minutes spent by \SP to complete analysis.} \label{tb:eff}
	\resizebox{\linewidth}{!}{
		\begin{tabular}{|l|l|l|l|l|l|l|l|l|}
			\hline
			\multicolumn{1}{|c|}{\textbf{$\mathit{\sigma}$}}                                         & \textbf{10\%} & \textbf{20\%} & \textbf{30\%} & \textbf{40\%} & \textbf{50\%} & \textbf{60\%} & \textbf{70\%} & \textbf{80\%} \\ \hline \hline
			\textbf{VF\_source}                                                        & 12            & 26            & 45            & 70            & 105           & 158           & 245           & 420           \\ \hline
			\textbf{VF\_oo7}                                                           & 12            & 26            & 45            & 70            & 105           & 158           & 245           & 420           \\ \hline
			\textbf{\begin{tabular}[c]{@{}l@{}}Analysis \\time\\ (Minutes)\end{tabular}} & 25.18         & 26.15         & 28.43         & 30.5          & 32.03         & 32.6          & 33.53         & 43.6          \\ \hline
	\end{tabular}}
\end{table}

For further evaluating the effectiveness of \SP, we design an experiment to check whether \SP can detect all potentially vulnerable code. We select the program {\tt bzip2} from SPECint CPU benchmark suite~\cite{henning2006spec}, and insert several vulnerable functions to the source code of {\tt bzip2}. The vulnerable functions are randomly chosen from the Spectre v1 variants suggested by Kocher~\cite{spectremitigations}. Assume that {\tt bzip2} has $P$ functions, we use $\mathit{\sigma}$ to represent the ratio of inserted vulnerable functions, where $ \mathit{\sigma = N / (N+P)}$ and $N$ is the number of inserted vulnerable functions. We use different values for $ \mathit{\sigma}$ to evaluate the effectiveness of \SP. The invocation to each vulnerable function is inserted in random locations of the {\tt bzip2} source code. Besides, each vulnerable function contains a taint source to guarantee the vulnerable code can be controlled by the attacker.

TABLE~\ref{tb:eff} presents the number of functions inserted in the source code of {\tt bzip2} (VF\_source), and the number of vulnerable functions identified by \SP (VF\_oo7) in the modified program with $\mathit{\sigma}$ ranging from 10\% to 80\%. Program {\tt bzip2} contains $P = 105$ functions, so we insert $\lceil \mathit{\sigma} \times 105 / (1-\mathit{\sigma}) \rceil$ vulnerable functions. For example, we insert 12 randomly picked vulnerable functions in the {\tt bzip2} code when $\mathit{\sigma} = 10\%$. As observed in TABLE~\ref{tb:eff}, our analysis can identify all the vulnerable code fragments over the varying range of $\sigma$ (i.e. [10\%, 80\%]). Moreover, as shown in TABLE~\ref{tb:eff}, the analysis time increases from 25.18 minutes with $\mathit{\sigma} = 10\%$ to 43.6 minutes with $\mathit{\sigma} = 80\%$. 
%

\begin{figure}[]
	\centering
	\includegraphics[width=1\linewidth]{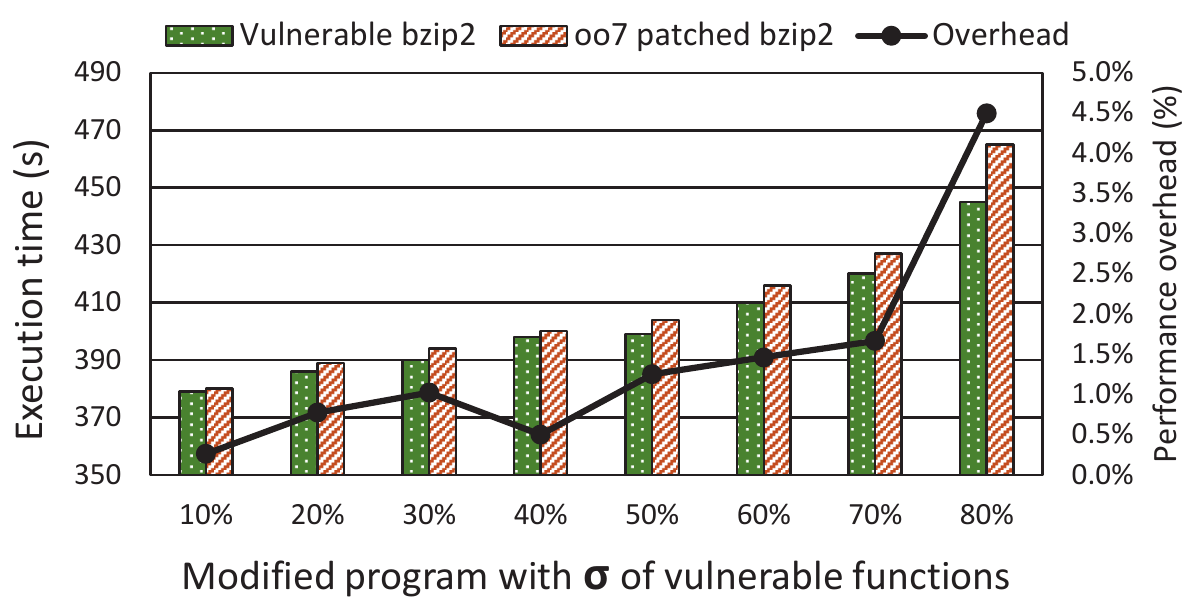}
	\caption{Execution time and performance overhead comparison between vulnerable {\tt bzip2} with ratio $\mathit{\sigma}$ vulnerable functions inserted and the patched program repaired by \SP.}
	\label{fig:eff}
\end{figure}

Fig.~\ref{fig:eff} show the execution time of the modified vulnerable {\tt bzip2} and the patched version by \SP along with the performance overhead introduced by patching. The absolute execution time in fig.~\ref{fig:eff} are shown by bars, and the performance overhead is shown by the line with markers. As shown in fig.~\ref{fig:eff}, the execution time is higher when $\mathit{\sigma}$ is increased, for example, the test with $\sigma=10\%$ can finish in 379 seconds, but the test with $ \mathit{\sigma=80\%}$ finished in 445 seconds. We use fence instructions to repair the program. We note that fence instructions introduce extra runtime overhead, as they prevent speculative execution. We can see from fig.~\ref{fig:eff} that the repaired program with $ \mathit{\sigma=10\%}$ takes only one second more than the vulnerable program, but the repaired program with $\mathit{\sigma=80\%}$ takes 20 seconds more than the vulnerable program. Moreover, the repaired program has negligible performance overhead. Specifically, the minimum overhead is 0.26\% (when $\sigma=10\%$) and the maximum overhead is about 4.49\% (when $\sigma=80\%$).

\subsection{Evaluation on Specint Benchmarks}
\label{sec:specfix-runtime-overhead}
\begin{figure*}[ht]
	\centering
	\includegraphics[width=0.95\linewidth]{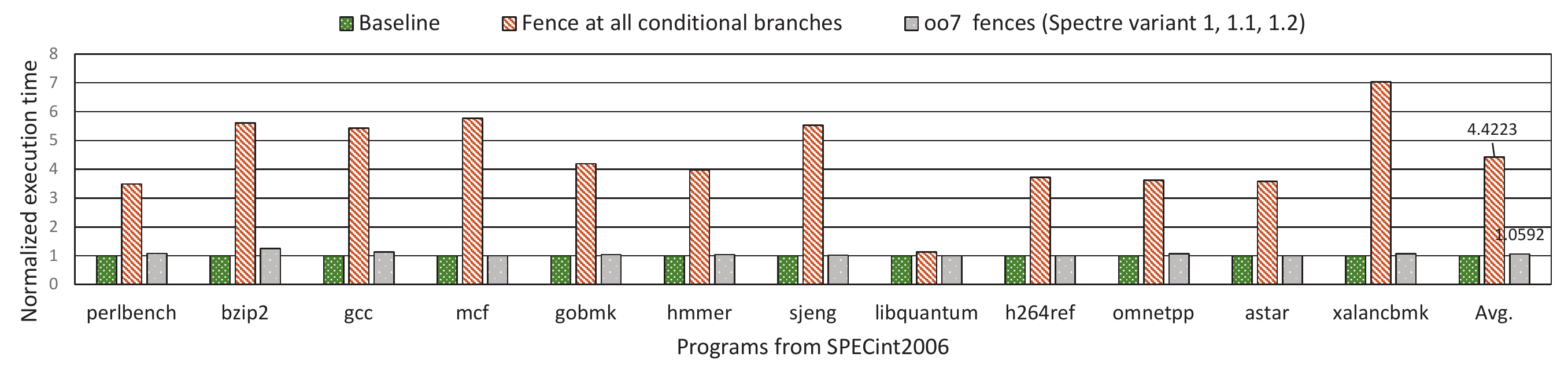}
	\caption{Runtime overhead due to protection against Spectre Variant 1, 1.1, 1.2 compared to original code in SPECint.
    {\em All} denotes the overhead from inserting fence at all conditional branches. The overheads from fences introduced by {\em oo7} to protect against Spectre variants 1, 1.1, 1.2 are plotted.}
	\label{fig:p2}
\end{figure*}


\begin{table*}[]
    \scriptsize
	\caption {Results for the detection of Spectre vulnerable code fragments in SPECint. \#$\langle TB \rangle$ denotes the tainted conditional branches detected by \SP. $\langle TB, RS \rangle$  satisfies Spectre Variant 1 condition, while $\langle TB, SW \rangle$ satisfies  Spectre Variant 1.1, 1.2 conditions as detected by \SP.} \label{tb:program}
	\resizebox{\linewidth}{!}{	
\begin{tabular}{|l|r|r|r|r|r|r|r|r|}
	\hline
	\textbf{Program}    & \textbf{Binary Size} &\textbf{LoC} & \textbf{\begin{tabular}[c]{@{}l@{}}Analysis \\ time(h|m)\end{tabular}} & \textbf{\begin{tabular}[c]{@{}l@{}}Repair\\ time (s)\end{tabular}} & \textbf{\begin{tabular}[c]{@{}l@{}}\#Conditional\\ branches\end{tabular}} & \textbf{\# $\langle TB \rangle$} & \textbf{\# $\langle TB, RS\rangle$} & \textbf{\#$\langle TB, SW\rangle$} \\ \hline \hline
	\textbf{perlbench}  & 1.2MB  &97,314              & 125h                                                                   & 5                                                                  & 21,972                                                                   & 60                                     & 18                                         & 5                                          \\ \hline
	\textbf{bzip2}      & 69 KB    &5,115            & 27.4m                                                                  & 1                                                                  & 942                                                                     & 102                                    & 81                                         & 5                                          \\ \hline
	\textbf{gcc}        & 3.6MB     &365,844           & $>$ 150h                                                                   & 11                                                                 & 59,614                                                                   & 103                                     & 8                                          & 2                                          \\ \hline
	\textbf{mcf}        & 23 KB    &1,370            & 3.5m                                                                    & 4                                                                  & 202                                                                     & 42                                      & 0                                          & 0                                          \\ \hline
	\textbf{gobmk}      & 3.9 MB   &154,170            & 19.2h                                                                  & 1                                                                  & 11,549                                                                   & 57                                     & 13                                          & 0                                          \\ \hline
	\textbf{hmmer}      & 319KB    &19,267            & 3.67h                                                                  & 1                                                                  & 4,468                                                                    & 49                                     & 15                                         & 0                                          \\ \hline
	\textbf{sjeng}      & 153 KB   &10,147            & 2.6h                                                                   & 1                                                                  & 2,146                                                                    & 18                                     & 3                                          & 0                                          \\ \hline
	\textbf{libquantum} & 51KB     &2,212            & 1.35h                                                                   & 1                                                                  & 444                                                                     & 30                                     & 0                                          & 0                                          \\ \hline
	\textbf{h264ref}    & 577 KB   &32,623            & 25.3h                                                                    & 1                                                                  & 6,743                                                                    & 16                                      & 0                                          & 0                                          \\ \hline
	\textbf{omnetpp}    & 768KB    &22,603            & 21h                                                                  & 2                                                                  & 4,812                                                                    & 90                                     & 26                                         & 3                                          \\ \hline
	\textbf{astar}      & 52 KB    &3,003         & 21.2m                                                                   & 1                                                                  & 541                                                                     & 19                                       & 0                                          & 0                                          \\ \hline
	\textbf{xalancbmk}  & 5.8MB    &186,997            & $>$ 150h                                                                   & 15                                                                 & 62,209                                                                   & 72                                       & 47                                          & 2                                          \\ \hline
\end{tabular}}
\end{table*}


\begin{table}[]
	\caption{Detected taint sources for programs from SPECint}
	\label{tb:taint_source}
	\begin{tabular}{l|l}
		\hline
		\textbf{Program}    & \textbf{Taint source list}                                                                                                                                                                         \\ \hline\hline
		perlbench  & \begin{tabular}[c]{@{}l@{}}getpid(), getgid(), getuid(), geteuid(), getegid(), \\ read(), fgetc(), getcwd(), getenv(), \\ gettimeofday(), fread()\end{tabular}           \\ \hline
		bzip2      & read()                                                                                                                                                                     \\ \hline
		gcc        & \begin{tabular}[c]{@{}l@{}}fread(), read(), getcwd(), getenv(), \\ \_IO\_getc()\end{tabular}                                                                        \\ \hline
		mcf        & fgets()                                                                                                                                               \\ \hline
		gobmk      & fgets(), getenv() \_IO\_getc()                                                                                                                       \\ \hline
		hmmer      & fread(), \_\_fread\_chk(), fgets(), getenv()                                                                                                                                         \\ \hline
		sjeng      & fgets(), \_IO\_getc()                                                                                                                                 \\ \hline
		libquantum & fgetc(), fread(), getenv()                                                                                                                                                  \\ \hline
		h264ref    & \begin{tabular}[c]{@{}l@{}} fread(), read(), \_\_isoc99\_fscanf(), \end{tabular}                                            \\ \hline
		omnetpp    & fgets(), getenv(), \_IO\_getc()                                                                                                                                \\ \hline
		astar      &read(), fscanf()                                                                                                                                                    \\ \hline
		xalancbmk  & getcwd(), fread(),  getenv()                                                                                                                                        \\ \hline \hline
	\end{tabular}
\end{table}

We use SPECint CPU benchmark suite~\cite{henning2006spec} to quantify the performance overhead of \SP protection mechanism as well as for evaluating the efficacy of our detection and repair. 

SPECint benchmark suite contains twelve programs in C and C++. TABLE~\ref{tb:program} outlines the salient features of these program: the binary size, analysis and repair time, the number of conditional branches, the number of tainted branches $ \mathit{TB}$, the number of $\langle \mathit{TB}, \mathit{RS} \rangle$ pairs as well as the number of $\langle \mathit{TB}, \mathit{SW} \rangle$ pairs. We run \SP on the programs in SPECint benchmark suite for at most 150 hours; only {\tt gcc} and {\em xalancbmk} do not complete in 150 hours. This is because {\tt gcc} and {\tt xalancbmk} have higher number of conditional branches.


In this set of experiments, we treat any APIs from the standard library that can import data to the program as potential taint sources. TABLE~\ref{tb:taint_source} lists the taint sources identified by \SP in SPECInt benchmark suite. Most of the programs in SPECint read the input from one or more files; so the file reading functions from standard C library are taint sources in these programs. For example, in {\em gcc} benchmark, {\tt fread()}, {\tt read()}, {\tt getcwd()}, {\tt getenv()} and {\tt \_IO\_getc()} from {\tt glibc} are used in the code and are marked as taint sources. 



We note that eight out of twelve programs from Specint benchmark exhibit the vulnerability pattern of Spectre variant 1 as evidenced by the presence of $\langle TB, RS \rangle$ pattern in {\tt perlbench, bzip2, gcc, gobmk, hmmer, sjeng, omnetpp, xalancbmk}. By looking for the $\langle TB, RS \rangle$ pattern, we conservatively assume the strictest security requirement of reading secret data. The subsequent mechanism to leak the secret data can vary with the most common mechanism being the cache side-channel with $\langle TB, RS, LS\rangle$ code pattern. Five of the benchmarks ({\tt perlbench, bzip2, gcd, omnetpp, xalancbmk}) are vulnerable to Spectre variant 1.1, 1.2 as evidenced by the presence of $\langle TB, SW \rangle$ pattern. The analysis time varies from 3.5 minutes ({\tt mcf}) to 150 hours ({\tt gcc}). The analysis time not only depends on the binary size but also the complexity of the program logic, more specifically, the number of branches, for example {\tt gcc} and {\tt xalancbmk} contain 59,614 and 62,209 branches in there binary. Since {\tt gcc} and {\tt xalancbmk} contain more branches than other programs, there are more execution paths to explore. Therefore our analysis times out after 150 hours. Our repair works on the assembly code and can complete in 15 seconds for these benchmarks.

We evaluate the runtime overhead due to fence insertion by executing each modified program ten times and report the average values. Fig.~\ref{fig:p2} shows the normalized execution time. The average performance overhead is 430\% when fences are inserted naively at both paths of all conditional branches. This is the safest strategy in the absence of an accurate program analyzer such as \SP. In contrast, \SP only inserts fences at detected conditional branches covering Spectre Variants 1, 1.1, 1.2 and hence incurs only 5.9\% overhead on an average. 


\subsection{Evaluation on interactive program} 
\begin{figure}[t]
	\centering
	\includegraphics[width=1\linewidth]{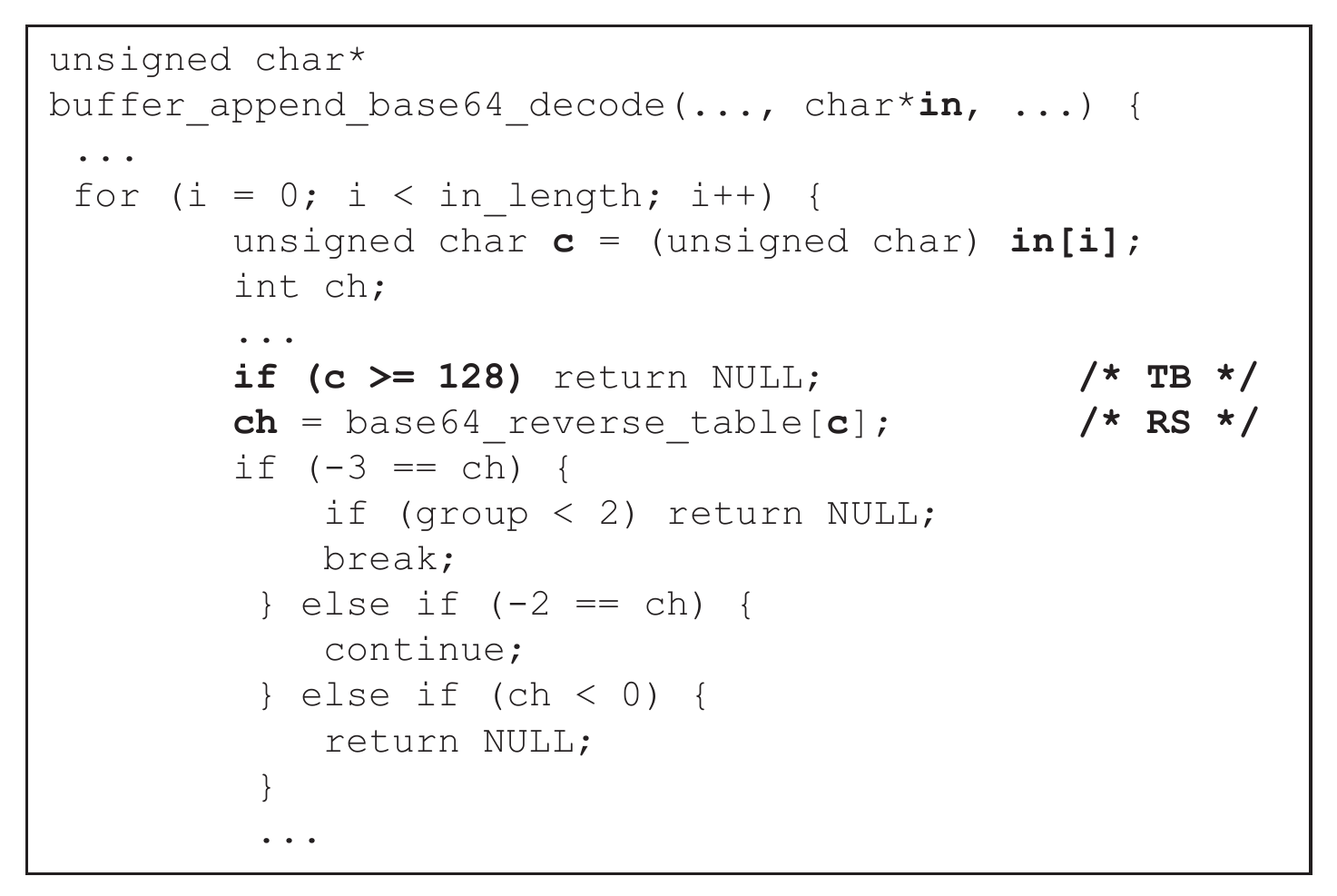}
	\caption{Potential Spectre vulnerability in {\tt base.c} within project
		{\em lighttpd}. {\tt in} is tainted from a taint source $recv()$. $\mathit{TB}$, $\mathit{RS}$ are highlighted.}
	\label{fig:lighttpd}
\end{figure}

{\tt lighttpd} is a lightweight web server which allows programs to interact with it by sending HTTP requests. We run \SP on {\tt lighttpd} to evaluate its effectiveness on interactive programs.  We identified nine taint sources in {\tt lighttpd}, detected 129 tainted branches and 40 Read Secret ($\mathit{RS}$). In the identified taint sources, {\tt recv()} is a critical interface that receives data from arbitrary clients through the network, which poses potential threats for data leakage in the server caused by deliberately constructed data from the attacker. Moreover, we evaluate the performance of repaired {\tt lighttpd} by sending 10000 HTTP requests to the {\tt lighttpd} server; the experiment shows our repaired code only slows down {\tt lighttpd} by 5.6\%.

Fig.~\ref{fig:lighttpd} shows a typical Spectre variant 1 vulnerability detected in function {\tt buffer\_append\_base64\_decode()} within file {\tt lighttp1.4/src/base64.c} of {\tt lighttpd} project. {\tt buffer\_append\_base64\_decode()} is called to decode the {\tt base64} string after {\tt recv()} receives data from the network. The code extracts a sequence of characters to variable {\tt c} from the input buffer {\tt in} and checks whether {\tt c} is less than 128 ($\mathit{TB}$). If $c < 128$, then the value of {\tt c} is used to index the table {\tt base64\_reverse\_table}. The attacker can first train the branch ``{\tt if (c >= 128)}" by using values of {\tt c} to be less than 128. Then the attacker can pass a value of {\tt c} to be 128. This results in accessing bytes outside of the array {\tt base64\_reverse\_table[]} being read to variable {\tt ch}. This happens when the branch ``{\tt if (c >= 128)}" is mispredicted for $c=128$ and during the speculative execution of the array access {\tt base64\_reverse\_table[]}. The attacker can infer whether the leaked data is equal to -3 by probing the 
cache line impacted by the following {\tt if} statement. Moreover, the type of {\tt c} is {\tt unsigned char} with a value range 0 to 255. Therefore this vulnerable code can at most leak 128 bytes outside of array {\tt base64\_reverse\_table[]}.

\subsection{Evaluation on various software projects}

\begin{table*}[]
\begin{center}
	\tiny
	\caption {Software projects used from Github and OSS-Fuzz and the detected Spectre v1 Vulnerabilities.} \label{tb:benchmark}
	\resizebox{\linewidth}{!}{
	\begin{tabular}{|l|l|l|l|l|l|l|l|l|l|l|l|}
		\hline
		\multicolumn{1}{|c|}{\multirow{2}{*}{\textbf{Project}}} & \multicolumn{1}{c|}{\multirow{2}{*}{\textbf{Project Description}}} & \multirow{2}{*}{\textbf{\begin{tabular}[c]{@{}l@{}}\# of \\ binaries\end{tabular}}} & \multirow{2}{*}{\textbf{\begin{tabular}[c]{@{}l@{}}Avg. \\ binary \\ size\end{tabular}}} & \multicolumn{4}{c|}{\textbf{Data-dependency}}                                                                                                                                                                                                                                                                                                                                         & \multicolumn{4}{c|}{\textbf{Program-dependency}}                                                                                                                                                                                                                                                                                                                                      \\ \cline{5-12} 
		\multicolumn{1}{|c|}{}                                  & \multicolumn{1}{c|}{}                                              &                                                                                     &                                                                                          & \textbf{\begin{tabular}[c]{@{}l@{}}\# of \\ vulnerable \\ binaries\end{tabular}} & \textbf{\begin{tabular}[c]{@{}l@{}}Avg. \# of \\ $\langle TB\rangle$ \\ per binary\end{tabular}} & \textbf{\begin{tabular}[c]{@{}l@{}}Avg. \# of \\ $\langle TB, RS\rangle$ \\ per binary\end{tabular}} & \textbf{\begin{tabular}[c]{@{}l@{}}Avg. \\ analysis \\ time(h)\end{tabular}} & \textbf{\begin{tabular}[c]{@{}l@{}}\# of \\ vulnerable \\ binaries\end{tabular}} & \textbf{\begin{tabular}[c]{@{}l@{}}Avg. \# of \\ $\langle TB\rangle$ \\ per binary\end{tabular}} & \textbf{\begin{tabular}[c]{@{}l@{}}Avg. \# of \\$\langle TB, RS\rangle$ \\ per binary\end{tabular}} & \textbf{\begin{tabular}[c]{@{}l@{}}Avg. \\ analysis \\ time(h)\end{tabular}} \\ \hline \hline
		\textbf{samba}                                          & SMB/CIFS networking protocol                                       & 230                                                                                 & 124.0KB                                                                                  & 62                                                                               & 16                                                                                                      & 9                                                                                                          & 0.76                                                                         & 91                                                                               & 32                                                                                                     & 46                                                                                                         & 1.2                                                                          \\ \hline
		\textbf{coreutils}                                      & GNU OS file, shell and text manipulation utilities                 & 114                                                                                 & 125.3KB                                                                                  & 78                                                                               & 14                                                                                                     & 4                                                                                                          & 0.22                                                                         & 84                                                                               & 83                                                                                                     & 57                                                                                                         & 1.3                                                                          \\ \hline
		\textbf{cups}                                           & Common UNIX Printing System                                        & 52                                                                                  & 134.3KB                                                                                  & 30                                                                               & 40                                                                                                     & 31                                                                                                         & 1.07                                                                         & 46                                                                               & 141                                                                                                    & 122                                                                                                        & 3.09                                                                         \\ \hline
		\textbf{freeradius}                                     & Popular open-source RADIUS server                                  & 47                                                                                  & 49.9KB                                                                                   & 18                                                                               & 13                                                                                                     & 24                                                                                                         & 0.21                                                                         & 25                                                                               & 82                                                                                                     & 25                                                                                                         & 0.45                                                                         \\ \hline
		\textbf{openldap}                                       & Lightweight Directory Access Protocol                              & 31                                                                                  & 1.3MB                                                                                    & 28                                                                               & 291                                                                                                    & 98                                                                                                         & 8.05                                                                         & 25                                                                               & 580                                                                                                    & 484                                                                                                        & 20.59                                                                        \\ \hline
		\textbf{openssh}                                        & Network utilities based on SSH protocol                            & 11                                                                                  & 791.9KB                                                                                  & 4                                                                                & 21                                                                                                     & 4                                                                                                          & 21.13                                                                        & 6                                                                                & 72                                                                                                     & 15                                                                                                         & $>$ 24                                                                           \\ \hline
		\textbf{xrdp}                                           & Remote desktop protocol (rdp) server                               & 10                                                                                  & 107.3KB                                                                                  & 5                                                                                & 23                                                                                                     & 2                                                                                                          & 0.59                                                                         & 5                                                                                & 48                                                                                                     & 22                                                                                                         & 3.04                                                                         \\ \hline
		\textbf{ppp}                                            & PPP daemon and associated utilities                                & 4                                                                                   & 322.0KB                                                                                  & 2                                                                                & 56                                                                                                     & 41                                                                                                         & 5.11                                                                         & 2                                                                                & 77                                                                                                     & 60                                                                                                         & 6.16                                                                         \\ \hline
		\textbf{dropbear}                                       & Small SSH server and client                                        & 4                                                                                   & 1.2MB                                                                                    & 2                                                                                & 148                                                                                                    & 20                                                                                                         & $>$ 24                                                                           & 2                                                                                & 172                                                                                                    & 44                                                                                                         & $>$ 24                                                                           \\ \hline
		\textbf{netdata}                                        & Distributed real-time performance monitoring                       & 2                                                                                   & 1.9MB                                                                                    & 2                                                                                & 109                                                                                                    & 45                                                                                                         & 12.05                                                                        & 2                                                                                & 198                                                                                                    & 175                                                                                                        & 12.41                                                                        \\ \hline
		\textbf{wget}                                           & Content retrieval from web servers                                 & 1                                                                                   & 937.2KB                                                                                  & 1                                                                                & 134                                                                                                    & 25                                                                                                         & 16.1                                                                         & 1                                                                                & 542                                                                                                    & 430                                                                                                        & $>$ 24                                                                           \\ \hline
		\textbf{darknet}                                        & Convolutional Neural Networks                                      & 1                                                                                   & 663.9KB                                                                                  & 1                                                                                & 76                                                                                                     & 42                                                                                                         & 3.17                                                                         & 1                                                                                & 183                                                                                                    & 195                                                                                                        & 6.8                                                                          \\ \hline
		\textbf{Total}                                          & -                                                                  & 507                                                                                 & -                                                                                        & 233                                                                              & -                                                                                                      & -                                                                                                          & -                                                                            & 290                                                                              & -                                                                                                      & -                                                                                                          & -                                                                            \\ \hline
	\end{tabular}}
\end{center}
\end{table*}

We observe that the detection of Spectre Variant 1 (as opposed to variants 1.1, 1.2) takes the longest time from our experiments with SPECint benchmarks. So we evaluate the scalability of Spectre variant 1 detection on real-world programs. In the evaluation, we select 507 binaries from OSS-fuzz repository of Google and other open-source projects; most of the selected binaries receive input data from the outside world through network interface. We apply \SP on the program binaries from selected projects as shown in TABLE~\ref{tb:benchmark}. The analysis are performed for at most 24 hours to detect potential Spectre variant 1 code snippets to demonstrate the scalability of our analysis.

The column "\# of vulnerable binaries" in TABLE~\ref{tb:benchmark} shows the number of program binaries in each project with potential vulnerabilities. We identify a program as vulnerable if it has at least one $\langle TB, RS\rangle$ pattern in the code that can potentially be exploited by the attacker to read secret data, i.e., we conservatively assume the strictest security requirement. As mentioned earlier, the subsequent mechanism to leak the secret data can vary with the most common mechanism being the cache side-channel with $\langle TB, RS, LS\rangle$ code pattern. For each project, we report the number of vulnerable programs, the average number of $\langle TB\rangle$ and the average number of $\langle TB, RS\rangle$ patterns under two different taint propagation strategies: data dependencies and program (data \& control) dependencies. For example, in project {\em samba}, out of total 230 programs, \SP detects 62 and 91 binaries as potential victims under data- and program-dependence taint propagations, respectively. Program-dependence based taint propagation identifies additional vulnerable code fragments compared to data-dependence only. Altogether, 233 (or 290) out of 507 programs are labeled as potential victims by \SP under data- (or program-) dependence taint propagation. TABLE~\ref{tb:benchmark} also show the analysis time in hours for detecting $\langle TB,RS\rangle$ patterns
using data dependencies, and program dependencies.

\begin{figure}[t]
	\centering
	\includegraphics[width=0.9\linewidth]{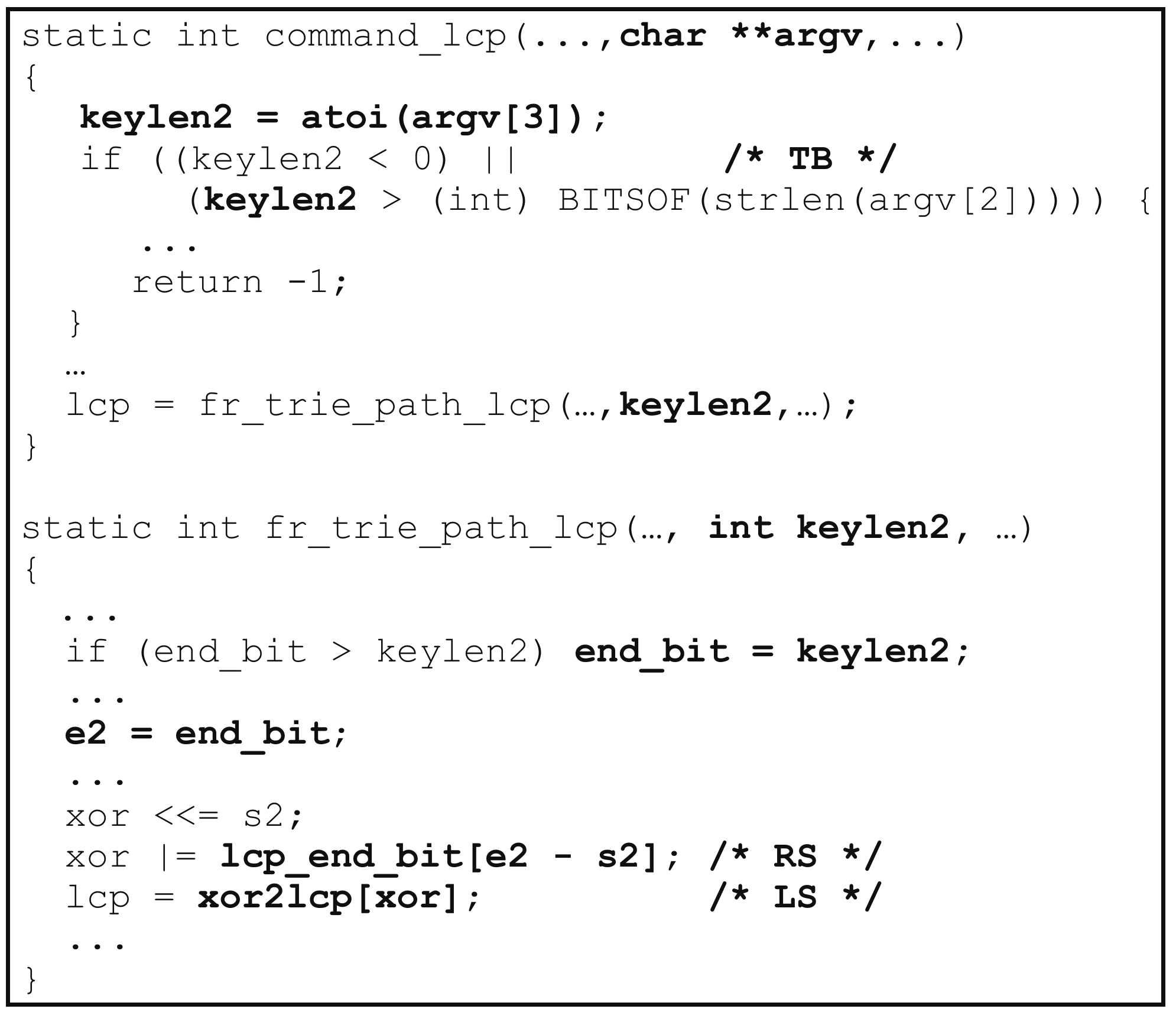}
	\caption{Potential Spectre vulnerability in {\tt trie.c} within project
	{\em freeradius}. {\tt argv} is tainted from a taint source $gets()$. The triplet $\langle \mathit{TB}, \mathit{RS}, \mathit{LS} \rangle$
	is highlighted.}
	\label{fig:spectre-real}
\vspace*{-0.2in}
\end{figure}

\paragraph*{\bf Potential Spectre Vulnerability in Large-scale Code}
We show one example of a Spectre vulnerability unearthed by \SP in
Fig.~\ref{fig:spectre-real}. This code snippet is identified by \SP in a program
({\tt src/lib/util/trie.c}) within the project {\em freeradius}. 
Note that \SP identifies the vulnerability at binary level. For the sake of exposition and brevity, we show only the portions corresponding
to the Spectre vulnerability pattern at source code level. As comments, we highlight the
code fragments detected as $\mathit{TB}$, $\mathit{RS}$ and
$\mathit{LS}$. The argument {\tt argv} is a tainted array read from an external file through the taint source {\tt gets()}. The conditional check in the function
{\tt command\_lcp} is therefore a tainted branch ($\mathit{TB}$).
Taint is propagated to the function
{\tt fr\_trie\_path\_lcp} via the parameter {\tt keylen2}.
Consequently, the array load {\tt lcp\_end\_bit} may use the value
of {\tt e2} (potentially controlled by the attacker
through {\tt argv[3]}) during speculative execution. This speculative
execution may take place due to the misprediction of the conditional
branch in {\tt command\_lcp} that reflects a bound check.
Finally, the array access {\tt xor2lcp} may reveal information out of the boundary of array {\tt lcp\_end\_bit[]} via cache side channel. Though the pattern $\langle TB,RS,LS \rangle$ is found in the wild by \SP, the vulnerable code fragment is executed only once at run-time, making it impossible for the adversary to poison the branch and launch an attack.
The distance between $\mathit{TB}$ and $\mathit{RS}$ is 145 Intel x86 instructions. The example illustrates that Spectre vulnerability in real-world  may span over multiple functions requiring inter-procedural program analysis.

\begin{figure}[h]
	\centering
	\includegraphics[width=1\linewidth]{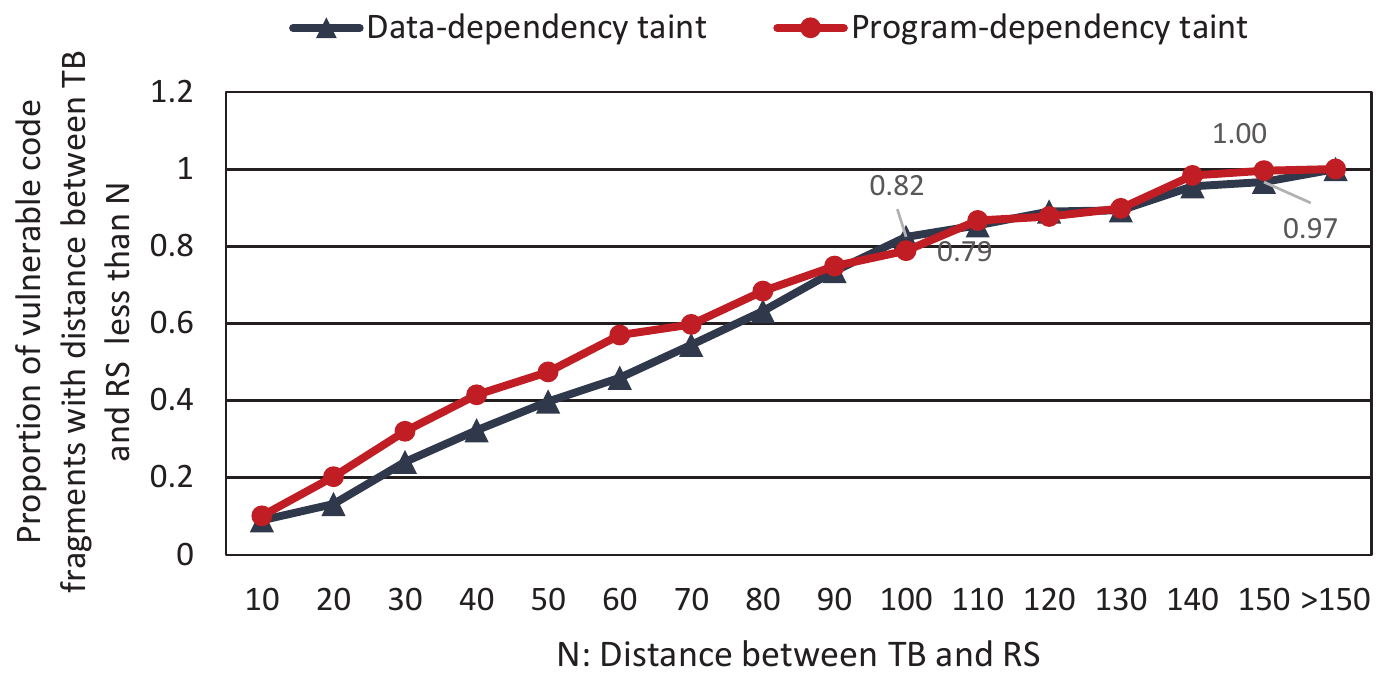}
	\caption{Cumulative distribution of distance (\#instructions) between TB and RS for binaries in TABLE~\ref{tb:benchmark} under data- and program dependence based taint propagation.}
	\label{fig:distance}
\end{figure}



\paragraph*{\bf Sensitivity to Speculative Execution Window size}
We set the speculative execution window size $SEW = 448$ as twice the effective ROB size in our platform. This is a conservative assumption to take care of micro-operations generated from the instructions and processed within the micro-architecture. We investigate the sensitivity of our analysis on $SEW$ value. Fig.~\ref{fig:distance} shows the distance in instructions between TB and RS ($\Delta(TB, RS)$) for vulnerable code fragments across all 507 binaries. The results show that 82\% and 79\% of the tainted memory accesses (RS) occur within 100 instructions from the tainted branch (TB) for data-and program-dependence based taint propagation, respectively.


\begin{figure}[ht]
	\centering
	\includegraphics[width=1\linewidth]{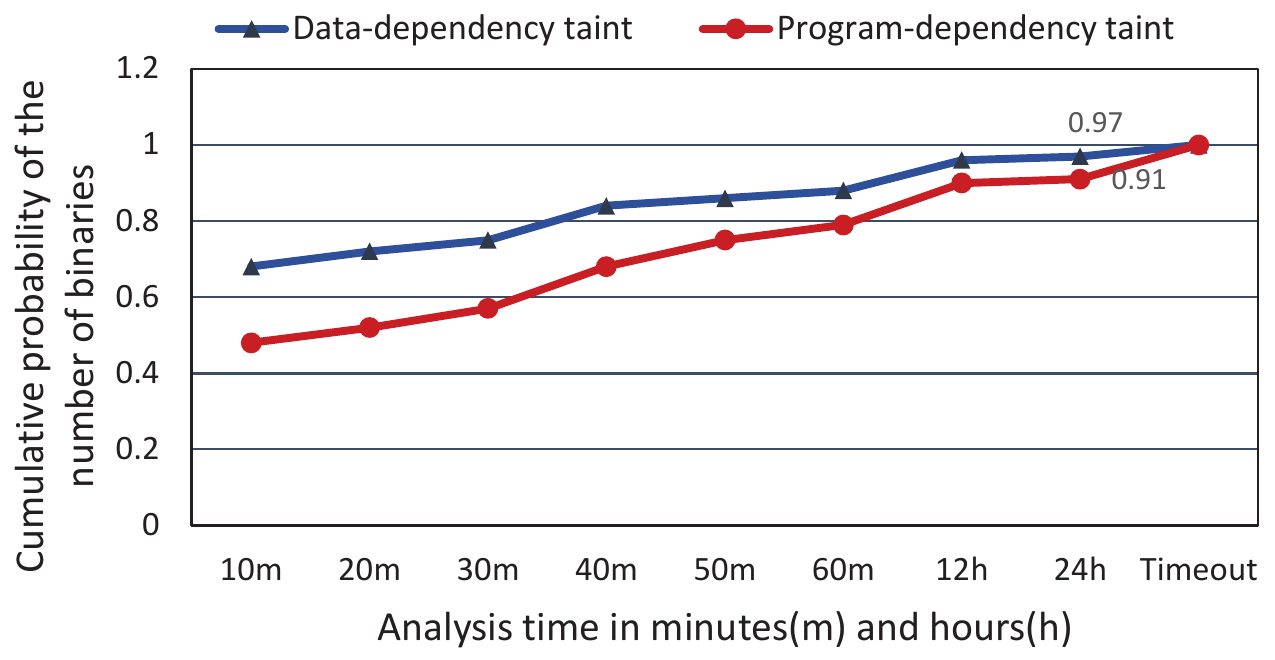}
	\caption{Cumulative distribution of analysis time for binaries in TABLE \ref{tb:benchmark} under data- and program dependence based taint propagation. }
	\label{fig:time}
\end{figure}

\paragraph*{\bf Analysis and Repair Time of \SP}
\label{sec:specfix-analysis-time}
The analysis time depends on the size and complexity of the binary. Fig.~\ref{fig:time} shows the distribution of analysis time across all the binaries. Under data-dependence based taint propagation, the analysis time is less than 20 minutes for 72\% of the binaries. Program-dependence based taint propagation increases the analysis time; still the analysis completes within 20 minutes for 52\% of the binaries. Only 3\% and 9\% of the analysis across 507 binaries did not complete in 24 hours. The repair time is minimal; for all of the 507 binaries, it is within $30$ seconds.

\paragraph*{\bf Quantitative Analysis of Vulnerabilities}
Our analysis shows that on an average only 7.3\% (variance 0.3\%) of conditional branches are tainted across 290 programs with at least one tainted branch. Moreover, 217 out of 507 binaries do not have any tainted branch at all.

Next we check the percentage of conditional branches in these program binaries that are tainted (TB) and are followed by tainted memory access (RS) within speculative execution window. If we want to ensure strict security requirements, then {\tt lfence} instruction should be inserted after all these tainted branches. On an average, our analysis shows only 3.72\% (variance 0.3\%) of conditional branches satisfy this criteria leading to very low overhead in fixing Spectre vulnerability.  

Finally, we check the percentage of conditional branches in these program binaries that are tainted (TB) and are followed by tainted memory access (RS) within speculative execution window and a subsequent tainted memory access (LS) to leak the data to the cache. This is denoted as {\em TB+RS+LS}. If we assume cache side-channel attack as the only mechanism to leak the secret brought into the cache, then \SP only needs to add {\tt lfence} instruction after these branches. On an average, only 2.32\% of conditional branches (variance 0.1\%) satisfy this criteria.  This strongly indicates that the performance overhead from inserting fences suggested by our technique will be low. However we could not collect the exact performance overhead for all of these 507 binaries because it will involve running each binary against many inputs and averaging the performance overhead across inputs. Furthermore, for some of the binaries such as {\tt coreutils}, a large set of inputs (each input being a command) is possible. For this reason, we meaured performance overheads on SPECint benchmarks instead, which is a standard benchmark suite with inputs specified for performance analysis. 

\paragraph*{\bf Higher vulnerable code fragment percentage in program-dependency enabled analysis. } As shown in TABLE~\ref{tb:benchmark}, \SP detects more vulnerable code fragments with program-dependency analysis (that considers both data dependency and control dependency) compared to only data-dependency analysis. For example, in {\tt samba} project, \SP detects on an average 16 tainted branches per binary when considering only data-dependency. In contrast, with program-dependency analysis, the number of tainted branches detected increases to 32 branches on average per binary. The reason is that if a conditional branch is tainted, all the instructions that have control-dependency with the conditional branch are also marked as tainted. Thus, analysis with program-dependency enabled inevitably causes over-tainting.
%


\section{Limitations of our approach}
\label{sec:threat}


\SP relies on BAP, which, in turn incorporates a taint analysis engine. The taint analysis statically interprets the code by unrolling the
loops up to a certain depth. In order to ensure that our approach does not introduce false negatives we need to pay attention to the following three issues.
\begin{itemize}
	\item
First of all,
optimistic loop unrolling may introduce false negatives (missing vulnerabilities) in \SP. However, with correct
or worst-case loop bounds  being supplied to BAP, such a limitation can be mitigated.
\item
Secondly, taint sources are provided
to the taint analysis engine, and if taint sources are under-specified then the taint analysis may not identify all the branches that can be controlled by the attacker. We thus conservatively assume all user inputs via console, file, and network as taint sources.
\item
Finally, the completeness of the control-flow extraction also plays a role to decide whether our analysis will introduce false negatives. If the branch targets of register indirect jumps are
not identified, the control flow graph extracted from the binary will not be complete, and as a result the taint analysis results may miss tainted branches. Thus, our approach always depends on the control flow graph being as complete as possible, in trying to
ensure that we do not have false negatives in our analysis.
\end{itemize}

\SP finds tainted memory accesses following a tainted conditional branch within a fixed speculation window. Incorrect setting of this speculation window size may lead to false positives (window size too big) or false negatives (window size too small). We conservatively set the window length to twice the size of the of the reorder buffer, as explained earlier.

\SP works on native code for program binaries. We have not investigated Spectre detection on interpreted code.

We also assume that all memory references in the victim code have been protected with appropriate
checks to prevent overflow and underflow  (e.g., array bound overflow). Thus, there does not exist any overflow/underflow  error in normal (i.e., non-speculative) execution traces.

The taint analysis capability gives \SP the flexibility to adapt to Spectre variants. We have discussed in detail (Section \ref{sec:overview}) the class of Spectre variants that we can handle and the ones that cannot be handled. In addition, new variants are constantly being found, we could face some variants in future that \SP cannot be adapted to handle.

The underlying technology used by \SP is forced execution, which is able to handle the obfuscated or self-modifying code~\cite{peng2014x}. Thus, \SP can potentially analyze obfuscated or self-modifying code. Moreover, code obfuscation and self-modifying code are common evasive techniques used by malware, 
while \SP focuses on identifying and repairing vulnerabilities in the general software instead of detecting malware.

\SP is  based on program analysis. As a result it can detect (and patch) vulnerable code patterns that can be exploited by malicious program inputs. It cannot detect scenarios where an external process affects micro-architectural states (e.g., flushing dynamic branch predictor) and thereby introduce vulnerabilities in another program (e.g., by exploiting the default static branch predictor as the dynamic branch predictor has been flushed). Such scenarios cannot be detected by our analysis.

\section{Discussion}
\label{sec:conclusion}

We have built \SP
for detecting Spectre vulnerabilities in
binary code and protecting against the attack with minimal overhead. Our approach is employed post-compilation on binary code to take
into account all the compiler optimizations. No change to the operating system or the processor is needed as the
approach proceeds by program analysis. We demonstrate that
systematic analysis is useful both for detecting Spectre
vulnerabilities and to repair them with minimal performance
overhead. 
Our work also provides an understanding of the class of Spectre attacks for which an analysis based mitigation may be suitable, and for which classes of attacks a system level solution is suitable. Our tool is publicly available from\\
{\tt \url{https://github.com/winter2020/oo7}}

\section{Acknowledgments}
This research was partially supported by a grant from the National Research Foundation, Prime Ministers Office, Singapore under its National Cybersecurity R\&D Program (TSUNAMi project, No. NRF2014NCRNCR001-21) and administered by the National Cybersecurity R\&D Directorate.




\balance
{\scriptsize
\bibliographystyle{IEEEtran}
\bibliography{ref/references}}


\vspace*{-0.2in}
\begin{IEEEbiography}[{\includegraphics[width=0.8in,height=1in,clip,keepaspectratio]{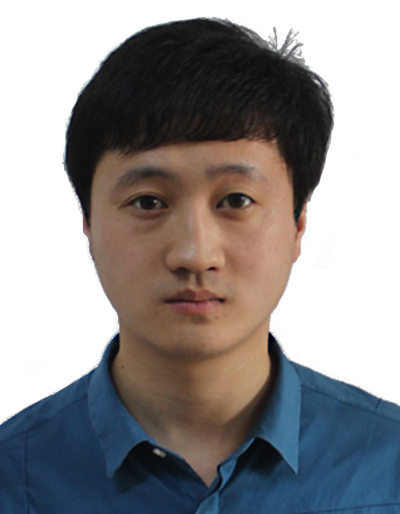}}]
		{Guanhua Wang} received the M.E degree in Computer science from Shandong University, Jinan, China in 2014, and he is currently pursuing the Ph.D degree with the Department of Computer Science, School of Computing, National University of Singapore. His research interests include CPU architecture, architecture security.
\end{IEEEbiography}

\vspace*{-0.3in}
\begin{IEEEbiography}[{\includegraphics[width=0.8in,height=1in,clip,keepaspectratio]{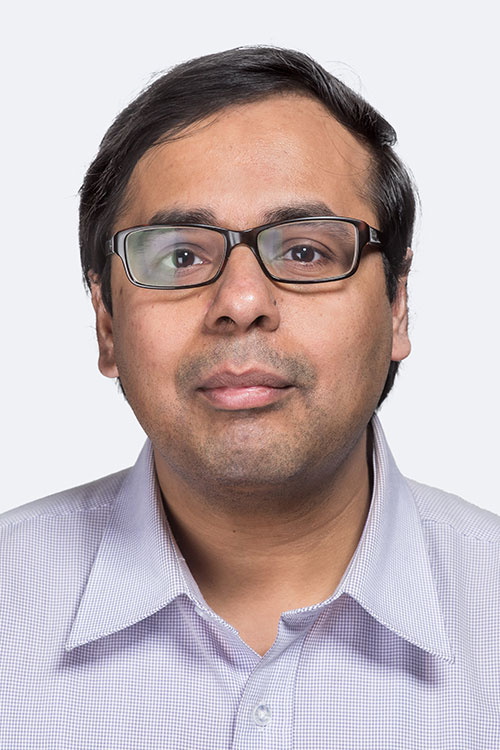}}]
	{Sudipta Chattopadhyay} received the Ph.D. degree
	in Computer science from the National University
	of Singapore, Singapore, in 2013.
	He is an Assistant Professor with the Information
	Systems Technology and Design Pillar, Singapore
	University of Technology and Design, Singapore.
	 His research interests include program analysis, embedded
	systems, and compilers.
\end{IEEEbiography}

\vspace*{-0.3in}
\begin{IEEEbiography}[{\includegraphics[width=0.8in,height=1in,clip,keepaspectratio]{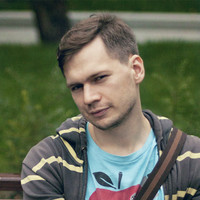}}]
	{Ivan Gotovchits} received his MS in Electrical Engineering from Moscow Technological University in 2006 and worked as an engineer developing mission-critical highly automated systems. He is a chief developer with CyLab, Carnegie Mellon University.
\end{IEEEbiography}

\vspace*{-0.3in}
\begin{IEEEbiography}[{\includegraphics[width=0.8in,height=1in,clip,keepaspectratio]{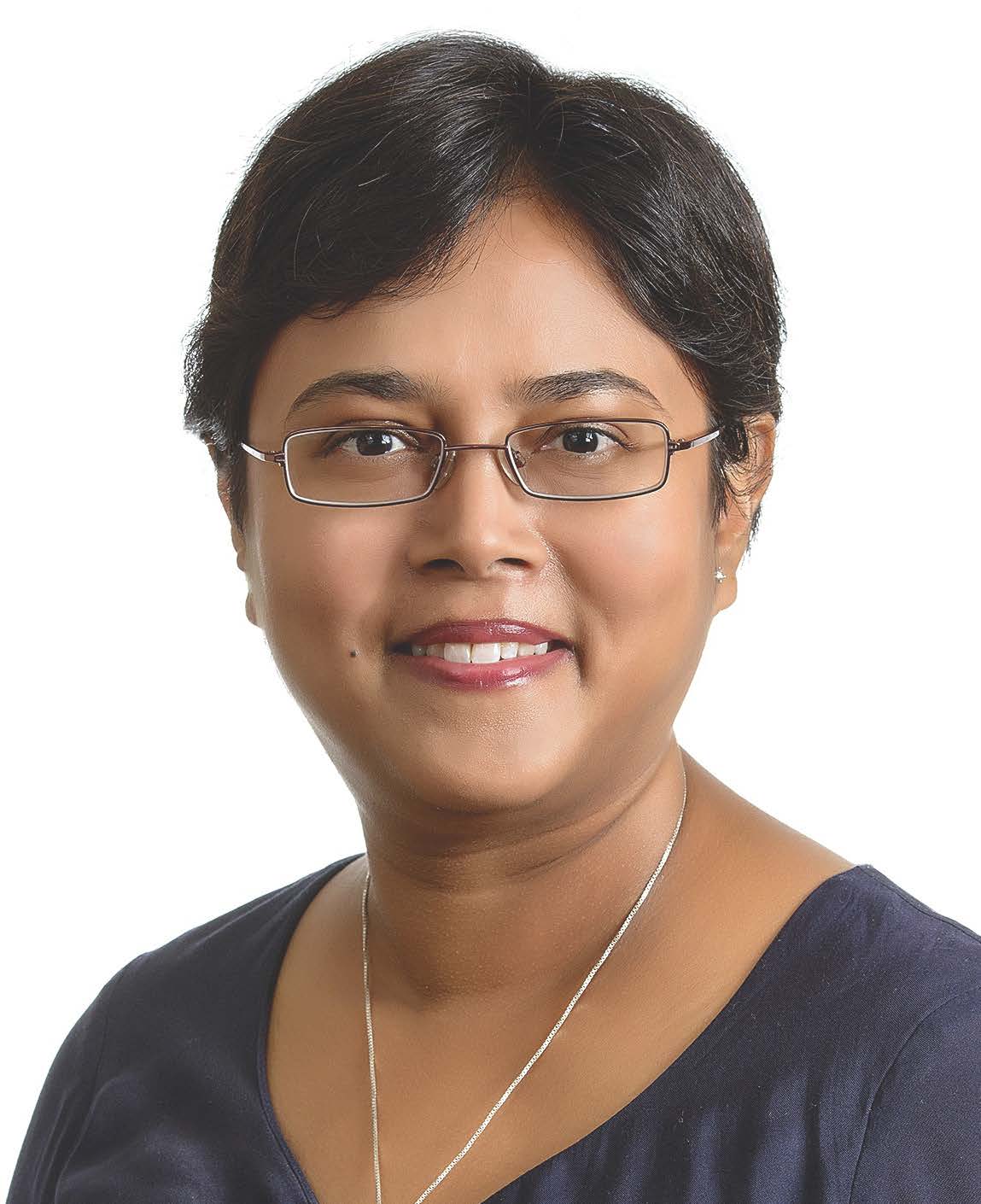}}]{Tulika Mitra}
	received a BE degree in Computer science from Jadavpur University, Kolkata, India, in 1995, an ME degree in Computer science from the Indian Institute of Science, Bengaluru, India, in 1997, and a PhD degree from the State  University of New York, Stony Brook, NY, USA, in 2000. She is currently a Professor of Computer science at the School of Computing, National University of Singapore, Singapore. Her research interests include the design automation of embedded real-time systems with particular emphasis on software timing analysis/optimizations, application-specific processors, energy-efficient computing, and heterogeneous computing.
\end{IEEEbiography}

\vspace*{-0.3in}
\begin{IEEEbiography}[{\includegraphics[width=0.8in,height=1in,clip,keepaspectratio]{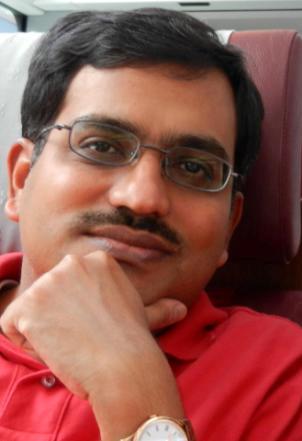}}]{Abhik Roychoudhury}
	received the Ph.D. degree in Computer science from the State University of New York at Stony Brook, Stony Brook, NY, USA, in 2000.
	He is a Professor of Computer science with the
	National University of Singapore, Singapore. His research interests include software testing and analysis, software
	security, and trust-worthy software construction.
	Dr. Roychoudhury is the Director of the National Satellite of Excellence in Trustworthy Software Systems, based in Singapore.
    He is also the Lead Principal Investigator of the Singapore
	Cyber-Security Consortium. He
	has served as an Associate Editor of the IEEE Transactions on Software Engineering from 2014 to 2018, and is currently serving
as an Associate Editor of the IEEE Transactions on Dependable and Secure Computing.
\end{IEEEbiography}

\end{document}